\documentclass[11pt,a4paper]{article}
\usepackage{jstyle, enumitem}
\usepackage{hyperref}
\usepackage{slashed,framed}
\usepackage{empheq}
\usetikzlibrary{calc,decorations.markings}

\author[a]{Jin-Beom BAE}
\author[b,c]{\quad Euihun JOUNG}
\author[d]{\quad Shailesh LAL}

\affiliation[a]{Scranton Honors Program, Ewha Womans University, Seoul 120-750, Korea}
\affiliation[b]{School of Physics and Astronomy, Seoul National University, Seoul 151-747, Korea}
\affiliation[c]{Gauge, Gravity \& Strings, Center for Theoretical Physics of the Universe, Institute for Basic Sciences, Daejeon 34047, Korea }
\affiliation[d]{LPTHE -- UMR 7589, UPMC Paris 06, Sorbonne Universit{\'e}s,  Paris 75005, France}

\emailAdd{kastalean4@gmail.com}
\emailAdd{euihun.joung@snu.ac.kr}
\emailAdd{shailesh@lpthe.jussieu.fr}

\begin{document}

\title{\centering On the Holography of Free Yang-Mills}

\abstract{We study the AdS$_5$/CFT$_4$ duality where the boundary CFT is free Yang-Mills theory with gauge group SU(N). At the planar level we use the spectrum and correlation functions of the boundary theory to explicate features of the bulk theory. Further, by computing the one-loop partition function of the bulk theory using the methods of arXiv:1603.05387, we argue that the bulk coupling constant should be shifted to $N^2$ from $N^2-1$.
Similar conclusions are reached by studying the dualities
 in thermal AdS$_5$ with \mt{S^1\times S^3} boundary.}

\maketitle

\section{Introduction}

Yang-Mills theory, with or without coupling to external matter, is known to admit a large-color expansion  strongly suggestive of a dual string description of the theory \cite{'tHooft:1973jz}. 
While a precise formulation of such a dual description of, say, quantum chromodynamics (QCD) remains lacking, string theory indicates that certain supersymmetric cousins of QCD such as \mt{\mathcal{N}=4} super-Yang-Mills in four dimensions are completely equivalent to string theory defined in Anti-de Sitter (AdS) spacetimes of one higher dimensionality \cite{Maldacena:1997re,Gubser:1998bc,Witten:1998qj}. 
Over the past two decades, this duality has yielded a number of important insights into otherwise intransigent problems such as the dynamics of strongly coupled quantum field theories (see \cite{Aharony:1999ti} for a review and further references).

At the same time, new AdS/CFT dualities have also been conjectured \cite{Sezgin:2002rt,Klebanov:2002ja} between certain non-supersymmetric CFTs and higher-spin (HS) theories in AdS.
A novel feature of these dualities is that
the relevant CFTs are in the vector --- 
rather than the adjoint --- representation
of the internal symmetry group, typically $U(N)$ or $O(N)$.
This drastically reduces the set of
single-trace operators in CFT, hence the spectrum of the bulk theory:
it only consists of fields in a single `Regge trajectory'\footnote{During this paper we will occasionally refer to fields that lie in `$n$-th Regge trajectories', by which
we mean the set of fields dual to the operators involving $n+1$ CFT fields. In this terminology, the higher-spin fields are said to lie on the `first Regge trajectory'.},
which are mostly massless HS fields.
The precise duality reads
that the free scalar/spinor CFT$_3$ in $U(N)$/$O(N)$ vector multiplet
corresponds to the type A/B non-minimal/minimal Vasiliev theory  in four dimensions \cite{Vasiliev:1990en}.
These AdS$_4$/CFT$_3$ dualities have higher dimensional counterparts
as well as the AdS$_3$/CFT$_2$ version.
In the latter case, the Prokushkin-Vasiliev theory \cite{Prokushkin:1998bq} in three dimensions is conjectured to be the dual of $W_N$ minimal model \cite{Gaberdiel:2010pz}.
About higher dimensions, the conjecture relates the any-$d$ non-minimal/minimal Vasiliev theory \cite{Vasiliev:2003ev}, based on a different type of oscillator, to the free scalar CFT in $U(N)/O(N)$ vector representation. 

These dualities, which we collectively refer to as HS/CFT dualities have been extensively investigated with a number of promising tests and developments. 
For instance, these dualities admit 
extensions of a critical deformation \cite{Klebanov:2002ja} as well as the one with a parity-violating structure \cite{Aharony:2011jz,Giombi:2011kc}.
The reader may consult the reviews \cite{Giombi:2012ms,Giombi:2016ejx} for further reading and references.
 Importantly for the present work, the matching of bulk and boundary partition functions at next-to-leading order 
has been studied in the series of works \cite{Giombi:2013fka, Giombi:2014iua, Giombi:2014yra} and \cite{Beccaria:2014xda,
Beccaria:2014zma,Beccaria:2014qea,Beccaria:2015vaa,Beccaria:2016tqy}.

In four boundary dimensions, conventional CFT can 
utilize not only scalar or spinor but also spin-one field.
In the papers \cite{Beccaria:2014xda,Beccaria:2014zma}, 
the authors studied the HS/CFT duality with
the free spin-one CFT transforming in the vector representation of $U(N)$ or $O(N)$.
Even though the corresponding bulk theory
--- which they referred as to ``Type C" --- is not yet explicitly constructed,
the authors could identify its field content 
and calculate the one-loop
partition functions 
for the cases of AdS$_5$ with $S^4$ boundary and \textit{thermal} AdS$_5$ with $S^1\times S^3$ boundary.

These developments naturally motivate us to examine 
whether the original expectation of \cite{'tHooft:1973jz} can be recast in some way in the modern AdS/CFT formulation. 
In particular, to elucidate the features of an AdS/CFT duality involving free Yang-Mills theory as the CFT. 
This subject has been explored on multiple fronts, ranging from a study of the thermodynamics of the CFT \cite{Sundborg:1999ue,Aharony:2003sx} to studies of the spectrum 
\cite{Gunaydin1998,Barabanschikov:2005ri,Newton:2008au}, but it would be fair to characterize our knowledge of this duality as preliminary, given our current ignorance of formulation of the stringy bulk theory leaving aside indications from general AdS/CFT considerations.

In this paper we shall confine ourselves to the following concrete questions: assuming that a dual formulation of free Yang-Mills with gauge group $SU(N)$ exists in AdS$_5$ in the planar limit of the gauge theory, is it possible to extend the duality to the next-to-leading order in the 't-Hooft expansion? 
More concretely, we attack the task
of evaluating one-loop partition function
of the Bulk Dual theory of free Yang Mills (BDYM).
The field content of BDYM can be in principle
identified by studying the single-trace operator spectrum
of free Yang-Mills,
which can be grouped into
different Regge trajectories.
The field content of the first Regge trajectory
actually coincides with that of minimal Type-C theory,
dual to $O(N)$ vector model of Maxwell theory,
and involves infinitely many massless HS gauge fields.
For higher Regge trajectories, finding the precise field content
might be feasible for the first few $n$, but 
quickly become technically prohibitive as $n$ grows.
This technical difficulty stands
as the main obstacle in studying an adjoint model 
holography.
In our previous work \cite{Bae:2016rgm} on the free 
scalar adjoint model holography,
we devised a new method 
--- which we henceforth refer to as the Character Integral Representation of the Zeta function (CIRZ)  ---
for the computation of bulk partition function
in order to bypass our ignorance of the explicit spectrum of the theory.
In the present work, we again employ the CIRZ method
in calculating one-loop partition function of BDYM.
Similarly to the scalar adjoint model case, 
we find the result  in a good agreement with the duality conjecture with the proviso that the `naive' dictionary between the bulk loop expansion parameter and the boundary large-color expansion parameter acquires a non-trivial shift.

A brief overview of this paper is as follows: 
In Section \ref{review},
we review some generalities about the AdS/CFT correspondence
with the focus on AdS$_5$/CFT$_4$, which is the case 
relevant to us.
We shall recollect some important facts about the representation theory of the conformal algebra $so(2,4)$, 
and general features of the bulk/boundary dictionary which we shall use for some specific free CFTs, including free Yang-Mills theory.
We conclude the section  with a review of the Character Integral Representation of the Zeta function (CIRZ) \cite{Bae:2016rgm}, which is the main technical tool 
of this paper.
Section \ref{typec} is devoted to a review of the Type-C duality and a reproduction of the one-loop vacuum energy,
now by using the CIRZ method. 
In Section \ref{freeym},
we move to the holography for the free Yang-Mills theory.
We first discuss some facts about the spectrum of single-trace operators as well as some general expectations about its AdS$_5$ dual that follow from representation theory. As an addendum, we also discuss briefly the higgsing of higher-spin symmetry that is expected to take place when a weak coupling is turned on for the Yang-Mills theory. 
With these inputs, in Section \ref{freeymtest},
we study the one-loop partition function of the
Bulk Dual theory to free Yang-Mills (BDYM) around  AdS$_5$ 
with $S^4$ boundary, by making use of the CIRZ method.
The result defines the AdS$_5$ vacuum energy of BDYM,
which is related to the $a$-anomaly coefficient of the free Yang-Mills.
Section \ref{sec:TAdS} is devoted to the case where
the background of BDYM is the \textit{thermal} AdS$_5$
with \mt{S^1\times S^3} boundary.
Here, the one-loop partition function gives
the Casimir energy of  thermal AdS$_5$.
We compute this quantity
both for the bulk dual of free adjoint scalar CFT as well as BDYM, contrasting them with each other.
As we shall see, a large part of our analysis for thermal AdS
goes through in any dimensions.
However, for the most part we shall confine ourselves to the case of $d=4$.
Finally, Section \ref{conclusion} contains our conclusion and additional 
discussions.

\section{A Few Prerequisites}\label{review}

In this work, we study the AdS$_5$ theory, dual to the  free $SU(N)$ Yang-Mills in four dimensions.  We will refer to this theory as BDYM (Bulk Dual theory to free Yang-Mills) henceforth. Compared to the free scalar model duality studied in the previous work \cite{Bae:2016rgm},
this model involves massless mixed-symmetry gauge fields in the bulk. These are precisely the fields that form the spectrum of the minimal Type-C duality of \cite{Beccaria:2014xda,Beccaria:2014zma}. For a better grasp of this theory, let us first review some essential ingredients for our study.
\subsection{UIRs and Characters of $so(2,4)$}
We begin with a review of the unitary irreducible representations (UIR) of the conformal/isometry algebra $so(2,4)$, 
which are carried by the perturbative spectrum of the theory. The lowest-weight representations $\cV(\D, (\ell_1,\pm \ell_2))$ of $so(2,4)$ are labelled by
 the quantum numbers  
 $\Delta$ and $(\ell_1,\pm\ell_2)$ of the lowest-weight state,
 which are the UIR labels of the subalgebra $so(2)\oplus so(4)$. 
 Since $so(4)\simeq su(2)\oplus su(2)$, the $so(4)$ label $(\ell_1,\pm \ell_2)$ can be
 translated into the $su(2)\oplus su(2)$ one $[j_\pm,j_\mp]$ as
 \be
 	\ell_1=j_++j_-\,,\qquad 
 	\ell_2=j_+-j_-\,.
\ee 
To avoid any confusions, we assume  $\ell_2\ge0$ and  $j_+\ge j_-$\,.
Under the parity transposition, $\ell_2$ flips its sign hence a parity-invariant theory
should include 
  $(\ell_1,\ell_2)_{\rm\sst PI}:=(\ell_1,+\ell_2)\oplus (\ell_1,-\ell_2)$
or equivalently $[j_+,j_-]_{\rm\sst PI}:=[j_+,j_-]\oplus [j_-,j_+]$ for $\ell_2\neq0$.
In the bosonic case where $\ell_1$ and $\ell_2$ are integers,
the parity-invariant representations
 can be realized as $so(4)$ tensors
with the index symmetry given by the Young diagram,
\be 
\parbox{95pt}{
	\begin{tikzpicture}
	\draw (0,0) rectangle (2.4,0.4);
	\draw (0,0.4) -- (0,0.8) -- (3.2,0.8) -- (3.2,0.4) -- (2.4,0.4);
	\node at (1.6,0.6){${\st \ell_1}$};
	\node at (1.2,0.2){${\st \ell_2}$};
	\end{tikzpicture}}\,.
\ee 
 Depending the value of $\D$, the $so(2,4)$ UIR
 with a given $(\ell_1,\pm\ell_2)$ fall into the three different classes \cite{mack1977all,Dolan:2005wy}.

\subsubsection*{Long Representation}

 When  $\Delta> \ell_1+2-\delta_{\ell_1\ell_2}$,
 the corresponding Verma module is unitary and irreducible:
 $\cD(\D,(\ell_1,\pm \ell_2))=\cV(\D,(\ell_1,\pm \ell_2))$.
 In CFT$_4$, it can be realized as a higher-spin operator,
\be
 	\cO^{a_1\cdots a_{\ell_1},b_1\cdots b_{\ell_2}}_\D(x)\,.
 	\label{Op}
\ee 
 In AdS$_5$, it is described by a massive higher-spin field,
 \be
	\varphi_{\m_1\cdots \mu_{\ell_1},\nu_1\cdots\nu_{\ell_2}}(z,x)\,, 
	\label{Massive F}
 \ee 
subject to the field equation given with the mass-squared
\cite{Metsaev:1995re,Metsaev:2004ee,Metsaev:2014sfa}, 
\be
	M^2=\frac{\D(\D-4)-\ell_1-\ell_2}{L^2}\,,
\ee
where $L$ is the radius of AdS$_5$\,.
When $\ell_2\neq0$\,, the operator \eqref{Op} and  the field \eqref{Massive F}
corresponding to UIR $\cD(\D,(\ell_1,\pm \ell_2))$
are restricted to satisfy the (anti-)self-duality condition depending on
the sign in front of $\ell_2$\,.
The models studied in this paper are all parity invariant, hence
involve the representation $\cD(\D,(\ell_1,\ell_2)_{\rm\sst PI})$
and the operator and field are not subject to the (anti-)self-duality condition.

The $so(2,4)$ character of this UIR is given by
\begin{equation}\label{long}
\chi_{\D,[j_+,j_-]}\!\left(q,x_+,x_-\right)
=q^{\Delta}\,P(q,x_+,x_-)\,\chi_{j_+}\!\left(x_+\right)\chi_{j_-}\!\left(x_-\right),
\end{equation}
where $\chi_j$ is the character of the spin-$j$ representation of $su(2)$ and takes the form
\begin{equation}
\chi_{j}\!\left(x\right)= \frac{x^{j+\frac12}-x^{-j-\frac12}}{x^{\frac12}-x^{-\frac12}}
=\frac{\sin(j+\frac12)\a}{\sin\frac{\a}2}
\qquad [x=e^{i\,\a}]\,,
\end{equation}
and $P(q,x_+,x_-)$ is given by 
\be
	P(q,x_+,x_-)
	=\frac1{\left(1-q\,x_+^{\frac12}\,x_-^{\frac12}\right)
	\left(1-q\,x_+^{-\frac12}\,x_-^{\frac12}\right)
	\left(1-q\,x_+^{\frac12}\,x_-^{-\frac12}\right)
	\left(1-q\,x_+^{-\frac12}\,x_-^{-\frac12}\right)}.
\ee
In the end, the character is a function of $q=e^{-\b}$ and $x_\pm=e^{i\,\a_\pm}$. Interpreted as partition function, $(\b,\a_+,\a_-)$ 
would correspond to the inverse temperature and two angular chemical potentials.

\subsubsection*{Semi-Short Representation}
\label{sec: semi-short}

When $\ell_1\neq \ell_2$ (that is, $j_-\neq 0$), 
the unitarity bound --- the smallest allowed conformal dimension for a unitary representation --- is $\Delta =\ell_1+2$ where
the Verma module develops an invariant subspace. 
The UIR is known as a semi-short representation and is given by the quotient $\cD(\ell_1+2,(\ell_1,\pm \ell_2))
=\cV(\ell_1+2,(\ell_1,\pm \ell_2))/\cV(\ell_1+3,(\ell_1-1,\pm \ell_2))$\,.
The semi-short representations relevant to the current work are the ones with
 the Young diagrams $(s,0)$ and $(s,2)_{\sst\rm PI}$.
 The former is the usual symmetric conserved current, whereas
 the latter can be realized as higher-spin operator $J^{a_1\cdots a_{s},b_1b_2}$ with the conservation condition,
\be
    \partial_{a_1}\,J^{a_1\cdots [a_{s-1} [a_s,b_1]b_2]}(x)=0\,,
    \label{conserv}
\ee
where the anti-symmetrization projects onto the Young diagram $(s-1,2)_{\rm\sst PI}$\,.
In AdS$_5$, these currents are dual to mixed-symmetry higher-spin fields
having the  gauge symmetry,
\be
	\delta\,\varphi_{\mu_1\cdots \mu_{s},
	\nu_1\n_2}(z,x)
	=\nabla_{(\mu_1}\,\varepsilon_{\mu_2\cdots \mu_{s)},
	\nu_1\n_2}(z,x)\,,
\ee
where $\nabla_\mu$ is the AdS covariant derivative.
In terms of parity-odd or even modes $(s,\pm 2)$,
the number degrees of freedom (DoF) of these fields 
coincides with that of symmetric spin-$s$ field, $2s+1$\,.
Even though $\cD(s+2,(s,\pm2))$ and $\cD(s+2,(s,0))$
have the same number of DoF, 
they are genuinely different representations in AdS$_5$.
In particular, they require different Goldstone modes to become massive. 
The Goldstone modes for the mixed-symmetry fields 
$\cD(s+2,(s,2)_{\rm\sst PI}))$ 
are nothing but the gauge modes $\cD(s+3,(s-1,2)_{\rm\sst PI})$,
which can be described by massive mixed-symmetry fields in AdS$_5$. 
The $so(2,4)$ character of these UIRs are given by the difference,
\ba\label{semishort}
&&
\chi_{\cD(s+2,(s,0))}=\chi_{s+2,(s,0)}-
\chi_{s+3,(s-1,0)}\,,\nn
&&
\chi_{\cD(s+2,(s,\pm 2))}=\chi_{s+2,(s,\pm2)}-
\chi_{s+3,(s-1,\pm2)}\,.
\ea

\subsubsection*{Short Representation} 

The last case is when $\ell_2=\ell_1=\ell$, or equivalently when $j_-$ vanishes.
The unitarity bound in this case lies on $\Delta = \ell+1=j_++1$\,. 
The corresponding lowest-weight module develops an invariant subspace
which itself contain again an invariant subspace.
The UIR is given by the coset,
$\cD(\ell+1,(\ell,\pm\ell))=\cV(\ell+1,(\ell,\pm\ell))/\cD(\ell+2,(\ell,\pm(\ell-1)))$
whereas $\cD(\ell+2,(\ell,\pm(\ell-1)))=\cV(\ell+2,(\ell,\pm(\ell-1)))/\cV(\ell+3,(\ell-1,\pm(\ell-1)))$\,.
The relevant short representation to this work
is the $\ell=1$ case.
In CFT$_4$, it is realized as the field strength operator $F^{ab}$
subject to the conservation condition, which is nothing but the equation of motion
for boundary spin-one field.
The character of this UIR is given by
\be\label{short}
\chi_{\cD(2,(1,\pm1))}=
\chi_{2,(1,\pm1)}
-\chi_{3,(1,0)}+\chi_{4,(0,0)}\,,
\ee
or equivalently (for the $(1,+1)$ case),
\ba
&&\chi_{\cD(2,[1,0])}(q,x_+,x_-)\nn
&&=\chi_{2,[1 ,0]}(q,x_+,x_-)-\chi_{3,[\frac12,\frac12]}(q,x_+,x_-)
+\chi_{4,[0,0]}(q,x_+,x_-)\nn
&&=q^{2}\,P(q,x_+,x_-)
\left[\chi_1(x_+)-q\,\chi_{\frac12}(x_+)\,\chi_{\frac12}(x_-)
+q^2\right]\nn
&&=e^{-\b}\,
\frac{2\left(\cosh\b\,\cos\frac{\a_+}2-\cos\frac{\a_-}2\right)\,
\cos\frac{\a_+}2
+\sinh\b\,\cos \a_+}
{2\left(\cosh\b-\cos\frac{\a_++\a_-}2\right)
\left(\cosh\b-\cos\frac{\a_+-\a_-}2\right)}\,.
\label{spin j char}
\ea
This character will play the key role in the subsequent analysis of this work.

\subsection{Tensor Product Decomposition}
\label{sec: NS}

In order to identify the spectrum of single-trace operators in a given model of free CFT,
one can rely on  group theoretical analysis.
The microscopic  information of CFT defines the UIR of the conformal field
and the symmetry of single-trace operators. The latter is governed by the global symmetry of CFT
and the representation  carried by the conformal field, such as
$O(N)$ vector or $SU(N)$ adjoint.
The entire set of independent single-trace operators is obtained by
decomposing tensor products of the UIR of the microscopic conformal field into UIRs of the conformal algebra $so(2,4)$. Typically, the former is short representation whereas the latter are long or semi-short.
The symmetry of single-trace operators determines
the maximum power and the symmetry (symmetric, cyclic, dihedral etc) of the products.
A convenient way to handle both of 
\begin{enumerate}[nolistsep,label=(\roman*)]
\item tensor products with different symmetries
\item decomposition into UIRs
\end{enumerate}
 is using the Lie algebra character.
Putting the issue (i) aside, in this section, we review a particularly efficient method for the decomposition (ii), developed in \cite{Newton:2008au}. 
However, since we will eventually bypass the task of identifying single-trace operator spectrum
by using the character
integral representation for the zeta function, devised in \cite{Bae:2016rgm}, 
this is not directly relevant to the computation at hand. The reader interested in the final results may safely skip this section for the later ones. 

Let us first consider, for illustrative purposes, a reducible representation $\cH$ of $su(2)$ symmetry.
Its character can be expanded as
\begin{equation}
\chi_{\cH}\!\left(x\right)=\sum_{j\in{\mathbb{N}/2}} N^{\cH}_j\,\chi_j\left(x\right),
\end{equation}
where $N^\cH_j$ is the multiplicity of the spin-$j$ representation in $\cH$, and in principle may also be zero. 
In this example, it is easy to check that the function $G_{\cH}\!\left(x\right)$ defined by
\begin{equation}
G_\cH\!\left(x\right)=\left\lbrace\left(1-{1\over x}\right)\chi_\cH\!\left(x\right)\right\rbrace_{x\geq 0}\,,
\end{equation}
satisfies the property that
\begin{equation}
G_\cH\!\left(x\right)= \sum_{j\in{\mathbb{N}/2}} N^{\cH}_j \,x^j\,,
\end{equation}
and hence serves as a generating function for the multiplicity of the spin-$j$ representation in $\cH$. Here the subscript $x\geq 0$ instructs is to expand the function enclosed in the braces in a series in $x$ and pick out the non-negative coefficients of $x$. 

Now, considering the case of $so(2,4)$,
a reducible representation $\cH$ ---
which would correspond to the spectrum of single-trace operators,
realized by certain tensor products of the microscopic conformal field  UIR ---
can be expanded as 
\be
\chi_{\cH}\!\left(q,x_+,x_-\right)= \sum_{\Delta,j_+,j_-}
N^\cH_{\cD(\Delta,[j_+,j_-])}\,
\chi_{\cD(\Delta,[j_+,j_-])}\!\left(q,x_+,x_-\right).
\label{UIR decomp}
\ee
The resulting UIRs $\cD(\D,[j_+,j_-])$ from the decomposition are in general
long or semi-short representations 
as they would correspond to single particle fields in AdS$_5$. 
As we have seen in Section \eqref{sec: semi-short}, the characters of semi-short representations
are simply given by the
difference of two long representation characters.
Hence, the character of the representation $\cH$ can be expanded as 
\be 
\chi_{\cH}\!\left(q,x_+,x_-\right)= \sum_{\Delta,j_+,j_-}
N^\cH_{\Delta,[j_+,j_-]}\,
\chi_{\Delta,[j_+,j_-]}\!\left(q,x_+,x_-\right),
\label{long decomp}
\ee
where $N^\cH_{\Delta,[j_+,j_-]}$ can be now negative integers.
When (semi-)short multiplets are present,
some readjustments are needed
for the translation of the multiplicities of Verma modules $N^\cH_{\Delta,[j_+,j_-]}$
 to those of conformal primaries $N^\cH_{\cD(\Delta,[j_+,j_-])}$ \cite{Newton:2008au}.  
 For example, the multiplicities 
 \mt{N_{s+2,[{s\over 2},{s\over 2}]}=n_1} and  \mt{N_{s+3,[{s-1\over 2},{s-1\over 2}]}=n_2} will lead to  \mt{N_{\cD(s+2,[{s\over 2},{s\over 2}])}=n_1} and $N_{\cD(s+3,[{s-1\over 2},{s-1\over 2}])}=n_2-n_1$.
Once the decomposition \eqref{long decomp} is known, it is straightforward to recover
the decomposition \eqref{UIR decomp}.
Similarly to the $su(2)$ example, we define
\begin{equation}\label{genfunc}
G_{\cH}\!\left(q,x_+,x_-\right)= \left\lbrace
\frac{\left(1-{1\over x_+}\right)\left(1-{1\over x_-}\right)
\chi_{\cH}\!\left(q,x_+,x_-\right)}{P(q,x_+,x_-)}\right\rbrace_{x_+\geq 0,\,x_-\geq 0},
\end{equation}
by filtering out the negative coefficients of $x_\pm$\,.
Then, one can show that the function $G_{\cH}$ actually generates the multiplicities $N^\cH_{\D,[j_+,j_-]}$\,:
\be
	G_{\cH}\!\left(q,x_+,x_-\right)=
	\sum_{\D,j_+,j_-}\,
	N^{\cH}_{\D,[j_+,j_-]}\,
	q^{\D}\,x_+^{j_+}\,x_-^{j_-}\,.
\ee
In Section \ref{sec: single trace}, we shall use this method to identify
the single-trace spectrum of 
the AdS$_5$ dual theory to the free Yang-Mills. However, it is more useful to consider a rewriting of the above generating function \eqref{genfunc} which yields the multiplicities of conformal primaries corresponding to a given twist \mt{\tau = \Delta -\ell_1}. Using the new variables $(\tilde x_1, x_2)$
where $\tilde x_1 = q\,x_1$ ($x_1 = 
{x}_+^{\frac12}\, 
{x}_-^{\frac12}$,
$x_2 =
{x}_+^{\frac12} 
{x}_-^{-\frac12}$), we obtain
\begin{equation}\label{genfuncx1x2}
\begin{split}
G\left(q,x_1,x_2\right) &= 
G\left(q,\tilde x_1/q,x_2\right)=
\sum_{\Delta,j_+,j_-} N_{\Delta,[j_+,j_-]}\,q^{\Delta-j_1}\,\tilde x_1^{j_+ + j_-}\,x_2^{j_+-j_-}\\ &= \sum_{\tau,\ell_1,\ell_2} N_{\tau,(\ell_1,\ell_2)}\,q^{\tau} \,\tilde x_1^{\ell_1}\, x_2^{\ell_2} = \sum_{\tau} q^{\tau}\, G^{\tau}\!\left(\tilde x_1,x_2\right),
\end{split}
\end{equation}
where $G^{\tau}$ counts the number of conformal primaries carrying a given twist $\tau$ but with otherwise arbitrary $SO(4)$ quantum numbers $\ell_1$ and $\ell_2$.

\subsection{Anomaly, Casimir Energy and AdS Vacuum Energy}

In this section, we provide a brief review  
of some essential features of the AdS/CFT correspondence that are useful for our present analysis. The reader is referred to \cite{Aharony:1999ti} for more details and references.
In particular, we review
how the one-loop partition function of the AdS theory
can be related to the anomaly and Casimir energy of the dual CFT.
 We shall also review some essential facts about the vacuum energy in AdS$_5$, and its computation at one loop by the zeta function method. This has already been reviewed extensively in \cite{Bae:2016rgm} and we refer the reader to that paper for more details and references. Subsequently, we shall also review the relevance of these computations to the holographic computation of the $a$-anomaly coefficient in CFT$_4$s \cite{Henningson:1998gx}. 

\subsubsection*{CFT Side}

Let us consider a CFT with action $S[\phi]$ where 
the (not necessarily scalar) field $\phi$ takes value in an $\mathsf{N}$-dimensional representation of the Lie algebra $SU(N)$. 
If $\phi$ carries the adjoint representation of $SU(N)$, $\mathsf{N}=N^2-1$. Let us now consider the free energy 
$F_{\rm\sst CFT}$ defined by
\begin{equation}
   \exp\!\left(- F_{\rm\sst CFT}\right) = \int\mathcal{D}\phi\, \exp\!\left(-S_{\rm\sst CFT}\left[\phi\right]\right).
\end{equation}
For a generic CFT this admits a large-$N$ expansion,
\begin{equation}
F_{\rm\sst CFT} = \mathsf{N}\,F^{\sst (0)}_{\sst\rm CFT} + F^{\sst (1)}_{\sst\rm CFT} + {1\over\mathsf{N}}\,F^{\sst (2)}_{\sst\rm CFT} +\cdots.
\end{equation}
For free CFTs,
the free energy $F^{\sst (0)}_{\rm\sst CFT}$ is one-loop exact and can be simply evaluated.

In even boundary dimensions, the free energy has a logarithmic UV divergence which is entirely fixed by the conformal anomaly as
\be 
    F^{\log}_{\rm\sst CFT}[g_{ab}]
    =\frac{\log\L_{\rm\sst CFT}}{(4\,\pi)^2}\,
    \int d^4x\sqrt{g}\left(
    a\,E_{\sst (4)}-c\,W^2\right)
    \label{F anom}
\ee
where $E_{\sst (4)}$ is the Euler characteristic 
density in four dimensions
and $W_{abcd}$ is Weyl tensor. 
Here, the coefficients $a$ and $c$ are respectively known 
as the $a$ and $c$ anomalies coefficients.

\paragraph{CFT on $S^4$}

When (Euclidean) AdS$_5$ has $S^4$ boundary, which is conformally flat,
the  $c$ anomaly identically vanishes and 
only $a$ anomaly gives the contribution to \eqref{F anom}.
Therefore, the free energy of a free CFT is given by
\be
    F_{\rm\sst CFT}
    =4\,a_{\rm\sst CFT}\,\log\L_{\rm\sst CFT}\,.
    \label{free F}
\ee 
In particular, the $a$-anomaly coefficient\footnote{In
the case of a generic free theory
involving $n_{0}$ scalars, $n_{\frac12}$ Dirac fermions and $n_1$ real vectors,
the $a$-anomaly coefficient is given by 
\begin{equation}
a_{\rm\sst CFT} = \frac{n_{0} +11\,n_{\frac12} + 62\,n_1}{360}\,.
\label{fuvdiv}
\end{equation}} for free Yang-Mills is
\be
    F_{\rm\sst free\,YM}
    =\mathsf{N}\,\frac{31}{45}\,\log\L_{\rm\sst CFT}\,.
    \label{free F YM}
\ee

\paragraph{CFT on $S^1\times S^3$}

The finite temperature CFT is defined by compactifying the (Euclidean) time direction to a circle $S^1$ with period $\beta$. 
We identify this period by inverse temperature. 
In this case,
the $a$ and $c$ anomalies all vanish hence
the log divergent term disappears.
Finally, the free energy of a free CFT is given by
\be
    F_{\rm\sst CFT}(\b)
    = \b\,E_{\rm\sst CFT}+
    \hat F_{\rm\sst CFT}(\b)\,,
    \label{free F th}
\ee
where $\hat F_{\rm\sst CFT}(\b)$ 
is the part vanishing in the low temperature ($\beta\to\infty$) limit.
Hence, in the limit, the free energy is dominated 
by the Casimir energy $E_{\rm\sst CFT}$\footnote{In
the case of a generic free theory
involving $n_{0}$ scalars, $n_\frac12$ Dirac fermions and $n_1$ real vectors,
the Casimir energy in $S^3$ is given by, see Table 2 of \cite{Beccaria:2014zma},
\begin{equation}\label{acoeffgeneral}
    E_{\rm\sst CFT} = \frac{n_{0} +{\frac{17}{4}}\,n_{\frac12} + 22\,n_{1}}{240}\,.
\end{equation}}
in $S^3$\,. In particular, the free energy of free Yang-Mills reads
\be
    F_{\rm\sst free\,YM}(\b)
    = \mathsf N\,\frac{11}{120}\,\b+
    \hat F_{\rm\sst free\,YM}(\b)\,,
\ee
where 
$\hat F_{\rm\sst free\,YM}(\b)$ 
is independent of $\mathsf{N}$ and given by
\begin{equation}
   \hat F_{\rm\sst free\,YM}(\b) =- \sum_{m=1}^\infty \frac1{m} \mathcal{Z}_{\rm\sst free\,YM}(m\,\b)\,,
\end{equation}
where $\mathcal{Z}_{\rm\sst free\,YM}(\b)$ is the
single-trace partition function of free Yang-Mills, which has been evaluated by P\'olya counting  in \cite{Sundborg:1999ue}.

\subsubsection*{AdS Side}

Now the formulation \cite{Gubser:1998bc,Witten:1998qj} of the AdS/CFT duality states that the quantity $F_{\rm\sst CFT}$ should be identified to the AdS quantity $\Gamma_{\text{AdS}}\left[h\right]$ given by
\begin{equation}
\exp\!\left(-\Gamma_{\sst\text{AdS}}\!\left[h\right]\right) = \prod_{\Delta,\ell,I}\int_{\varphi|_{\partial\text{AdS}}=h} \mathcal{D}\varphi^I_{\Delta,\ell}\,\exp\!\left(-{1\over g}\,S_{\sst\text{AdS}}\!\left[\phi\right]\right),
\end{equation}
evaluated at $h=0$. The subscript $\varphi|_{\partial\text{AdS}}=h$ of the path integral indicates that the fields obey the Dirichlet-like boundary conditions
$\varphi^{I}_{\Delta,\ell}\sim z^{\Delta}\,h^{I}_{\Delta,\ell}$
as $z$ tends to zero, which is the location of the AdS boundary in the Poincar\'{e} patch. Now we decompose the field $\phi^{I}_{\Delta,\ell}$ as 
$\varphi^{I}_{\Delta,\ell} = \varphi^{I}_{\Delta,\ell}\!\left(h\right) +\pi^{I}_{\Delta,\ell}$\,,
where $\varphi^{I}_{\Delta,\ell}\left(h\right)$ is the unique field configuration solving the classical equations of motion with the Dirichlet boundary conditions, and $\pi^{I}_{\Delta,\ell}$ denotes quantum fluctuations of the field $\varphi^{I}_{\Delta,\ell}$. 
In that case, $\Gamma_{\sst\rm AdS}\!\left[h\right]$ admits the loop expansion,
\begin{equation}
\Gamma_{\sst\text{AdS}}\!\left[h\right] = {1\over g}\,\Gamma^{\sst (0)}_{\sst\text{AdS}}\!\left[h\right] + \Gamma^{\sst (1)}_{\sst\text{AdS}}\!\left[h\right]+ g\,\Gamma^{\sst (2)}_{\sst\text{AdS}}\!\left[h\right]+\cdots\,.
\end{equation}
As mentioned previously, the AdS/CFT conjecture states that $\Gamma_{\sst\text{AdS}}\!\left[0\right]=F_{\sst\rm CFT}$. 
It turns out that the fairly innocuous property
that $F_{\rm\sst CFT}$ is free of $1/\mathsf{N}$ corrections
imposes fairly non-trivial constraints on the bulk theory. 
In particular, we expect the large-$N$ expansion on the boundary to be related to the loop expansion in the bulk. 
If we identify the dimensionless loop counting parameter $g$ in the bulk
with the boundary large-$N$ expansion parameter $1/\mathsf{N}$, then the absence
of the subleading contributions in the latter parameter implies that all higher-loop corrections to $\Gamma_{\sst\text{AdS}}$ must also vanish, that is, all the bubble diagrams in AdS must sum up to zero at every order in the loop expansion. In this paper we shall test this criterion for the case of the one-loop correction to $\Gamma_{\sst\text{AdS}}\!\left[0\right]$, given by $\Gamma^{\sst (1)}_{\sst\text{AdS}}\!\left[0\right]\equiv \Gamma^{\sst (1)}_{\sst\text{AdS}}$. This is given by
\begin{equation}\label{gamma1gen}
\exp\!\left(-\Gamma^{\sst (1)}_{\sst\text{AdS}}\right)=\prod_{\Delta,\ell,I}\int\,\mathcal{D}\pi_{\Delta,\ell}^I\,\exp\!\left(-\frac1g\,S_2\left[\pi\right]\right),
\end{equation}
where the quadratic action $S_2\left[\pi\right]$ is simply the sum of the quadratic actions for the various fluctuation fields $\pi_{\Delta,\bm\ell}^I$. This fact has the following important consequence. We can evaluate the vacuum energy piecewise, by computing it for every individual field with given $so(2,4)$ quantum numbers $\Delta,\,\bm\ell$ and then summing over all fields in the spectrum of the theory.
This is carried out through the \textit{zeta function} $\zeta_{\Delta,\bm\ell}\left(z\right)$ associated with 
$\cD(\D,\bm\ell)$ as 
\begin{equation}
\Gamma^{\sst (1)\,\rm ren}_{\sst\text{AdS}} = -{1\over 2}\,\sum_{\D,\bm\ell,I}\,N_{\Delta,\bm\ell}\,\zeta'_{\Delta,\bm\ell}\!\left(0\right),
\end{equation}
where $N_{\Delta,\bm\ell}$ is the number of fields carrying the $so(2,4)$ quantum numbers $\Delta,\bm\ell$.
\paragraph{AdS with $S^4$ boundary}

We shall provide explicit expressions for $\zeta'_{\Delta,\ell}\!\left(0\right)$ in Section \ref{cirz} but for the moment, let us note that the vacuum energy is proportional to the volume of AdS$_5$, i.e.
$\Gamma^{\sst (1)\,\rm ren}_{\sst\text{AdS}} =\pi^{-2} 
\,\text{Vol}_{\sst\text{AdS}_5}\,\gamma_{\rm\sst AdS}\,.$
This may be seen from the general expression \eqref{gamma1gen} and the homogeneity of AdS$_5$. 
The volume of AdS$_5$ is infinite, causing 
the volume factor to diverge, whereas the constant $\gamma_{\rm\sst AdS}$ 
is finite and depends on the AdS theory.
The volume divergence of AdS$_5$ is regulated by putting a radial cutoff at a large value $R$ of the AdS radial coordinate and we find that the regulated volume of AdS$_5$ is given by 
 $\text{Vol}_{\sst\text{AdS}_5} = \pi^2\,\log R$ \cite{Diaz:2007an}.
As a result, the vacuum energy takes the form
\begin{equation}
\Gamma^{\sst (1)\,\rm ren}_{\text{AdS}} = \log R\,\gamma_{\rm\sst AdS}\,,
\end{equation}
which will be eventually related to the 
conformal $a$-anomaly coefficient of the boundary theory
in Section \ref{freeymtest}

\paragraph{Thermal AdS with $S^1\times S^3$ boundary}

In the case of \textit{thermal} AdS, hereafter referred to as TAdS, which is the space obtained when the time direction in global AdS undergoes a periodic identification with period $\beta$,
the one-loop quantity $\Gamma^{\sst (1)\,\rm ren}_{\rm\sst TAdS}$
takes the form 
\begin{equation}
     \Gamma^{\sst (1)\,\rm ren}_{\rm\sst TAdS}(\b)= 
     \beta\,\cE_{\rm\sst TAdS} + 
     \hat\cF_{\rm\sst TAdS}(\b)\,,
     \label{TAdS F}
\end{equation}
where $\b\,\cE_{\rm\sst  TAdS}$ is the contribution 
proportional to the volume of thermal AdS, 
hence linear in $\beta$.
On the other hand, $\hat\cF_{\rm\sst TAdS}(\b)$ is the part
vanishing in the $\beta\to\infty$ limit. 
This quantity was computed in \cite{Giombi:2008vd,David:2009xg,Gopakumar:2011qs,Gaberdiel:2010ar,Gupta:2012he} for various spin fields in AdS. It will turn out that this contribution matches on the bulk and boundary side by construction.
However, the other term $\cE_{\rm\sst TAdS}$ will be of significance to us. 
It is the Casimir energy of the spatial section of thermal AdS
and will be eventually related to the boundary Casimir energy $E_{\rm\sst CFT}$ \cite{Giombi:2014yra}. 
We continue the discussion in Section \ref{sec:TAdS}, and compare the case of Yang-Mills theory with the free scalar $SU(N)$ adjoint model.

\subsection{Character Integral Representation of Zeta Function (CIRZ)}\label{cirz}

This section is a review of the formalism devised in \cite{Bae:2016rgm} to compute 
the one-loop  vacuum energy of a AdS theory with Hilbert space $\cH$\,.
In principle to calculate the full zeta function of the theory, we need to know first the field content of the AdS theory, that is,
the multiplicities $N^{\cH}_{\cD(\Delta,[j_+,j_-]}$:
\begin{equation}
\mathcal{H}=
\bigoplus_{\Delta,j_+,j_-}\,N^{\cH}_{\cD(\Delta,[j_+,j_-])}\,
\mathcal{D}\!\left(\Delta,[j_{+},j_{-}]\right),
\end{equation}
 as it is the sum of the zeta functions of individual fields:
 \be
 	\zeta_\cH(z)
 	=\sum_{\Delta,j_+,j_-}\,N^{\cH}_{\cD(\Delta,[j_+,j_-])}\,
 	\zeta_{\cD(\Delta,[j_+,j_-])}(z)\,.
 \ee
The perturbative spectrum, namely the field content, of AdS theory 
can be obtained by tensor product decomposition of the conformal field UIR,
which in turn amounts to decomposing the character for the whole theory, $\chi_{\cH}$,  into the UIR characters
$\chi_{\cD(\D,[j_+,j_-])}$ as in \eqref{UIR decomp}.
 The new method allows us to compute the zeta function
directly from the full character $\chi_{\cH}$ without needing to expand it into $\chi_{\cD(\D,[j_+,j_-])}$'s.
\begin{center}
	\begin{tikzpicture}
	\draw (0,0) rectangle (3.6,0.7);
	\node at (1.8,0.35){Full Character $\chi_{\cH}$};
	\draw (0,-1.2) rectangle (3.6,-0.5);
	\node at (1.8,-0.85){Spectrum};
	\draw (0,-2.4) rectangle (3.6,-1.7);
	\node at (1.8,-2.05){Zeta Function};
	\draw [ultra thick, red,->]  (-0.15,0.35) arc [radius=0.5, start angle=90, end angle = 270];
	\draw [ultra thick, red,->]  (-0.15,-1.05) arc [radius=0.5, start angle=90, end angle = 270];
	\draw [ultra thick, blue,->]  (3.75,0.35) arc [radius=1.4, start angle=60, end angle = -60];
	\node [red] at (-2,-0.85){Usual Method};
	\node [blue] at (6.2,-0.65){Character Integral};
	\node [blue] at (6.2,-1.05){Representation};
	\end{tikzpicture}
\end{center}
Here, let us only quote the final result of \cite{Bae:2016rgm}.
 The zeta function of the theory with Hilbert space $\cH$ is 
 given by the sum of three terms,
\begin{equation}
\zeta_{\mathcal{H}}\!\left(z\right):=\zeta_{\mathcal{H}|1}\!\left(z\right)+\zeta_{\mathcal{H}|2}\!\left(z\right)+\zeta_{\mathcal{H}|3}\!\left(z\right),
\end{equation}
where the $\zeta_{\mathcal{H}|n}$ are the Mellin transforms,
\begin{equation}\label{zetahn}
\frac{\Gamma(z)\,\zeta_{\mathcal{H}|n}(z)}{\log R} =\int_0^\infty {\left(\beta\over 2\right)^{2\left(z-1-n\right)}\over\Gamma\left(z-n\right)}f_{\mathcal{H}|n}\left(\beta\right),
\end{equation}
of the functions $f_{\mathcal{H}|n}$ defined through the character $\chi_{\cH}$ by
\begin{equation}\label{fhndef}
\begin{split}
f_{\mathcal{H}|2}(\b)&= {\sinh^4{\tfrac\beta2}\over 2}\,\chi_{\mathcal{H}}\left(\beta,0,0\right),\\
f_{\mathcal{H}|1}(\b)&= \sinh^2{\tfrac\beta2}\left[{\sinh^2{\tfrac\beta2}\over 3}-1-\sinh^2{\tfrac\beta2}\left(\partial_{\alpha_1}^2 +\partial_{\alpha_2}^2\right)\right]\chi_{\mathcal{H}}\left(\beta,\alpha_{+},\alpha_{-}\right)\bigg|_{\alpha_{\pm}=0},\\
f_{\mathcal{H}|0}(\b)&=\left[1 + {\sinh^2\tfrac\beta2\left(3-\sinh^2\tfrac\beta2\right)\over 3} \left(\partial_{\alpha_1}^2 +\partial_{\alpha_2}^2\right)\right.\\&\qquad\left. -{\sinh^4\tfrac\beta2\over 3}\left(\partial_{\alpha_1}^4-12\,\partial_{\alpha_1}^2\partial_{\alpha_2}^2+\partial_{\alpha_2}^4\right)\right]\chi_{\mathcal{H}}\left(\beta,\alpha_{+},\alpha_{-}\right)\bigg|_{\alpha_{\pm}=0}.
\end{split}
\end{equation}
In this manner, the zeta function associated with 
quadratic fluctuations of a given content $\cH$ of fields on AdS$_5$ 
may be written in terms of the character of the conformal algebra $so(2,4)$ associated with $\cH$. This formula enables us to bypass the explicit identification of the spectrum of single-trace operators and 
directly compute the zeta function, hence the one-loop vacuum energy of the theory about the AdS background. 
This is the key technical method which we shall use in this paper.

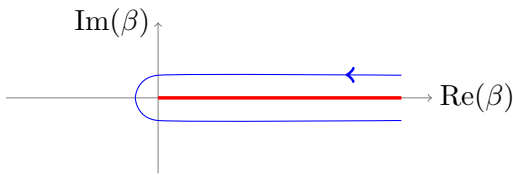
\begin{figure}[h]
\centering
\begin{tikzpicture}
\draw [help lines,->] (-2,0) -- (3.6,0);
\draw [help lines,->] (0,-1) -- (0,1);
\node at (4.2,0){Re$(\b)$};
\node at (-0.6,1) {Im$(\b)$};
\draw [line width =1.3pt, red] (0,0) to (3.2,0);
\draw [blue,
   decoration={markings,
  mark=at position 0.2 with {\arrow[line width=1.2pt]{>}}
  }
  ,postaction={decorate}]
  plot [smooth, tension=0.3]
  coordinates {(3.2,0.3) (0,0.3) (-0.3,0)};
  \draw [blue, postaction={decorate}]
  plot [smooth, tension=0.3]
  coordinates {(-0.3,0) (0,-0.3)  (3.2,-0.3)};
\end{tikzpicture}
\caption{Integration contour for the zeta function}
\label{fig: ct1}
\end{figure}

We now turn to discussing about the $\beta$ integrals given in \eqref{zetahn}. 
Using the Taylor expansions of the characters 
in $q=e^{-\beta}$, the $e^{-\beta\left(\Delta+m\right)}$ contribution in the integrand of \eqref{zetahn} leads to
\begin{equation}
\int_0^\infty d\beta\,
\frac{\left(\beta\over 2\right)^{2\left(z-1-n\right)}}{\Gamma\!\left(z-n\right)}\,
e^{-\beta\left(\Delta+m\right)}=4^{-z+n+1}\left(\Delta +m\right)^{-2z+2n+1}\,
\frac{\Gamma\!\left(2z-2n-1\right)}{\Gamma\!\left(z-n\right)}\,.
\end{equation}
Then it is easy to see that   the small $z$ behavior of the integral \eqref{zetahn}
is free of a singularity:
\begin{equation}\label{finteg}
\int_0^\infty {\left(\beta\over 2\right)^{2\left(z-1-n\right)}\over\Gamma\left(z-n\right)} f_{\mathcal{H}|n}\left(\beta\right) = -2\,\gamma_{\mathcal{H}|n} +\mathcal{O}\!\left(z\right),
\end{equation}
where $\gamma_{\cH|n}$ is a constant. 
From this we observe two things. Firstly, the UV divergence of the vacuum energy, corresponding to 
$\zeta_\cH\!\left(0\right)$ is universally absent in AdS$_5$. 
This is a well known fact about odd dimensions, having to do with the absence of integral powers (in particular $t^0$) in the short time $t$ expansion of the heat kernel in odd dimensions.
Secondly the finite part of the vacuum energy, controlled by $\gamma_{\mathcal{H}|n}$, is entirely captured by the divergence about the neighborhood of $\beta=0$. For  any finite field content $\cH$, the function $f_{\mathcal{H}|n}\left(\beta\right)$ has no singularities on the positive real axis except for the pole at $\beta=0$. 
Hence, for a sufficiently large $z$, the integral appearing in \eqref{finteg} may be recast into the contour integral
\begin{equation}\label{finteg2}
{i\over 2\sin\left(2\pi z\right)}\oint_C d\beta {\left(\beta\over 2\right)^{2\left(z-1-n\right)}\over\Gamma\left(z-n\right)} f_{\mathcal{H}|n}\left(\beta\right),
\end{equation}
where $C$ is the contour shown in Figure \ref{fig: ct1}. 
The advantage of the representation \eqref{finteg2} is that the above contour integral is well-defined for any value of $z$. In particular, the product of $\sin\left(2\pi z\right)\Gamma\left(z-n\right)$ has a smooth limit as $z$ approaches zero. Also, when we set $z$ to zero, the integrand becomes free of the branch cut and the contour may be shrunk to a small circle around $\beta=0$ and the value of the integral is controlled entirely by the residue of the function $f_{\mathcal{H}|n}$ at $\beta=0$. In particular,
\begin{equation}
\gamma_{\mathcal{H}|n} = -\left(-4\right)^n n! \oint {d\beta\over 2\pi i}\,{f_{\mathcal{H}|n}(\b)\over\beta^{2\left(n+1\right)}}\,.
\label{ct int}
\end{equation}
The above residues may easily be calculated from picking the $\beta^{2n+1}$ term in the small $\beta$ expansion of the functions $f_{\mathcal{H}|n}(\b)$ and the one-loop vacuum energy found to be the sum
\begin{equation}
\Gamma^{\rm\sst (1)\,\text{ren}}_\mathcal{H}=\log R\left(\gamma_{\mathcal{H}|2}+\gamma_{\mathcal{H}|1}+\gamma_{\mathcal{H}|0}\right).
\label{vac ener}
\end{equation}
To recapitulate, 
for any spectrum $\cH$ whose $f_{\cH|n}$ do not have any singularity 
on the positive real axis of $\b$,
both methods of evaluating the contour integral \eqref{ct int} and extracting 
$\beta^{2n+1}$ coefficient from $f_{\mathcal{H}|n}(\b)$ 
give the same answer. 
However, it is no more the case for  the theories which exhibit a Hagedorn transition, such as adjoint models,
since they possess branch points in the complex $\beta$ plane. 
As a result, in such a theory, the contour integral
would give an \textit{a priori} different result to that obtained from 
the $\beta^{2n+1}$  coefficient.
Let us emphasize that this ambiguity of choosing a proper prescription 
for the vacuum energy is not a mathematical or technical one, but a physical or conceptual one. 
For the moment, we do not find any clear physical guideline which give a preference
on one  than the other. 
Among the options, the one making use of the $\beta^{2n+1}$ coefficient
is computationally the most accessible to implement, 
and it is what we shall focus most closely on. 
But, let us remind again that the contour integral would in general
give a different result in the adjoint models we shall consider,
and indeed in the case of free scalar adjoint models.

\section{Holography for Type-C Higher-Spin Theory Revisited}\label{typec}

Before moving to the holography of free Yang-Mills, let us first consider
its vectorial counterpart,
which has been studied in \cite{Beccaria:2014xda,Beccaria:2014zma}.
The boundary CFT of the model is the $N$ complex or real Maxwell fields in the vector representation of $U(N)$ or $O(N)$,
and the AdS dual is a higher-spin field theory containing both symmetric and mixed-symmetry gauge fields. 
The latter AdS higher-spin theory is referred in \cite{Beccaria:2014zma} as to type C by analogy with
type A and B HS theories dual to free scalar and spin-half fermion CFTs.
We revisit this model as it can be viewed as the gauge or massless sector of the AdS theory dual to free Yang-Mills. 
More precisely, the set of `single-trace' operators in the vector model CFT is in one-to-one correspondence with the 
Yang-Mills single-trace operators involving two  curvatures.
This is very similar to how the type-A higher-spin theory can
be viewed as the gauge sector of the AdS dual of the free scalar adjoint model \cite{Bae:2016rgm}.

With this in mind, we turn to computing the vacuum energy
of the type-C HS theory, which has been first evaluated 
in \cite{Beccaria:2014xda}
and related to the anomaly $a$-coefficient of the boundary theory.
Even though the physical quantity we calculate is not new, 
our calculation method is so: 
in \cite{Beccaria:2014xda}, the total vacuum energy has been obtained
as the infinite sum of vacuum energies of each AdS fields with the damping factor $e^{-\e\,(s+\frac12)}$. 
In the type A model case, 
the latter regularization proved to give the same result
as the one obtained in the regularization where we first sum over the field content then remove the UV cut-off (which amounts to taking $z\to0$ limit).
However, finding the proper damping factor is not a systematic problem
in general and may become practically ambiguous in a case where there is no expected result and other regularization method is not available.
In the following, we compute the zeta function of the type C  higher-spin theory by using the Character Integral Representation 
of Zeta function (CIRZ), which is based on
the regularization scheme, `first sum over spectrum, then remove the regulator'.

\subsection{Vector Model Maxwell Theory and Its Single-Trace Operators}

The boundary CFT is based on Maxwell field, whose $so(2,4)$ UIR
corresponds to
\be 
	\cS_1=\cD(2,(1,1)_{\rm\sst PI})
	=\cD(2,(1,1))\oplus \cD(2,(1,-1))=\cD(2,[1,0])\oplus\cD(2,[0,1])\,,
\ee
which is the parity invariant combination of spin-one \emph{doubleton}
 --- which is also referred to as \emph{singleton}.
Combining the character \eqref{spin j char} with its $\cD(2,(1,-1))$ counterpart,
we obtain the character of parity invariant spin-one doubleton as
\be
\chi^{\phantom{g}}_{\cS_1}(\b,\a_+,\a_-)=e^{-\b}\,
\frac{e^\b\left(\cos^2\frac{\a_+}2+\cos^2\frac{\a_-}2\right)
-2\,\cos\frac{\a_+}2\,\cos\frac{\a_-}2-\sinh\b}
{\left(\cosh\b-\cos\frac{\a_++\a_-}2\right)\left(\cosh\b-\cos\frac{\a_+-\a_-}2\right)}\,.
\ee
In our analysis, this character plays the key role.
Like in type A and B models, there are two versions 
in the type C model holography: 
the duality between the $U(N)$ CFT and the
non-minimal type C higher-spin theory
and the duality between the $O(N)$ CFT and the minimal theory.

\subsubsection*{U(N)/Non-Minimal Model}

In the case of $U(N)$ model, the Maxwell field is complex, hence
the CFT is given by
\be
    S_{\rm CFT}=\int d^4x\,\sum_{i=1}^N \bar F_{i\,ab}\,F_i^{ab}\,.
\ee
All $U(N)$-invariant single-trace operators are bilinear in $\bar F_{i\,ab}$ and $F_{i\,ab}$\,.
The spectrum of such operator can be obtained by decomposing
\be
	\cH_\textrm{C,non-min}=\cS_1\otimes \cS_1\,,
\ee
into $so(2,4)$ UIR. This has been already carried out in \cite{Dolan:2005wy,Beccaria:2014xda,Beccaria:2014zma},
and the decomposition reads
\ba\label{j1spec}
	\cH_\textrm{C,non-min}\eq
	2\,\cD(4,(0,0))\oplus
	\cD(4,(1,1)_{\rm\sst PI})\oplus
	\cD(4,(2,2)_{\rm\sst PI})\nn 
	&&\oplus\,2
	\bigoplus_{s=2}^\infty 
	\cD(s+2,(s,0))\oplus
	\bigoplus_{s=3}^\infty
	\cD(s+2,(s,2)_{\rm\sst PI})\,.
\ea
The UIRs in the first line are long hence 
non-conserved currents on the boundary \cite{Beccaria:2014xda},
\be\label{j1currents1}
    \cO'=\bar F_{ab}\,{}^*F^{ab}\,,
    \qquad
    \cO=\bar F_{ab}\,F^{ab}\,,
    \qquad 
    \cO_{ab}=\bar F_{c[a}\,F_{b]}{}^c\,,
    \qquad 
    \cO_{ab}^{cd}=\bar F_{(a}{}^{(c}\,F_{b)}{}^{d)}\,,
\ee
dual to massive fields in AdS.
Here, the summation over internal $U(N)$ index should be understood
and $^*$ denotes the Hodge dual.
The UIRs in the second line are semi-short,
that is, conserved currents on the boundary dual to massless fields in AdS.
The symmetric conserved currents are of the form,
\be\label{j1currents2}
    J_{a_1\cdots a_s}
    = \bar F_{(a_1|b}\,\overset{\leftrightarrow}{\partial}_{a_2}\,
    \cdots\,\overset{\leftrightarrow}{\partial}_{a_{s-1}}\,
    F_{a_s)}{}^{b}
    +{}^*\bar F_{(a_1|b}\,\overset{\leftrightarrow}{\partial}_{a_2}\,
    \cdots\,\overset{\leftrightarrow}{\partial}_{a_{s-1}}\,
    {}^*F_{a_s)}{}^{b}
    -({\rm trace})\,,
\ee
where (trace) indicates the improvement term necessary to render
the current traceless.
The other symmetric conserved currents $J'_{a_1\cdots a_s}$
have the same form as above but $F_{i\,ab}$ 
and ${}^*F_{i\,ab}$ interchanged.
The precise form of these currents including the (trace) part and 
the mixed-symmetry conserved currents 
$J_{a_1\cdots a_s,b_1b_2}$
are given in \cite{Anselmi:1999bb}.

\subsubsection*{O(N)/Minimal Model}

When the Maxwell fields are real, $\bar A_{i\,\m}=A_{i\,\m}$,
then the $U(N)$ symmetry reduces to $O(N)$
and the single-trace operator spectrum
can be obtained by decomposing
the symmetrized tensor product,
\be
	\cH_{\rm C,min}=\cS_1\otimes_{\rm\sst sym} \cS_1\,.
\ee
The symmetrization projects out a part of single trace
operators and leaves
\ba \label{cyclicj1spec}
	\cH_{\rm C,min}\eq
	2\,\cD(4,(0,0))\oplus
	\cD(4,(2,2)_{\rm\sst PI})\nn 
	&&\oplus
	\bigoplus_{s=2}^\infty 
	 \cD(s+2,(s,0))\oplus \bigoplus_{s=4,6,\ldots} \cD(s+2,(s,2)_{\rm\sst PI})\,.
\ea
The projected operators are $\cO_{ab}$ and $J_{a_1\cdots a_s}'$
and $J_{a_1\cdots a_{2n+1},b_1b_2}$\,.
As opposed to the $U(N)$ case, the $O(N)$ spectrum
contains only one copy of symmetric currents/fields:
in particular, stress tensor $T^{\m\n}$
is the only rank-two current on the boundary,
or equivalently graviton is the only massless spin-two field in AdS.

\subsection{Vacuum Energy of Type-C Higher-Spin Theory}

We now revisit the computation of the 
vacuum energy in the (non-)minimal type C higher-spin theory in AdS$_5$. As mentioned before, 
we compute the zeta function (hence, the vacuum energy)
using the CIRZ method
rather than summing individual vacuum energies
with an ad hoc regularization prescription.

\subsubsection*{U(N)/Non-Minimal Model}

Let us first consider the $U(N)$/non-minimal model case.
In our method, all we need to know is the character of entire field content,
which is simply given by
\be 
	\chi^{\phantom{g}}_{\textrm{C,non-min}}(\b,\a_+,\a_-)=\chi^{\phantom{g}}_{\cS_1}(\b,\a_+,\a_-)^2\,.
\ee
Then, using the definitions \eqref{fhndef}, we obtain
\begin{equation}
\begin{split}
f^{\phantom{g}}_{\text{C,non-min}|2}(\b) &= \frac{e^{-2\,\beta } \left(3-e^{-\beta }\right)^2}{8 \left(1-e^{-\beta }\right)^2}\,,\\
f^{\phantom{g}}_{\text{C,non-min}|1}(\b) &= \frac{e^{-2 \beta } 
\left(3-e^{-\beta }\right)
\left(27-19\,e^{-\b}+17\,e^{-2\,\b}-e^{-3\,\b}\right)}{12 \left(1-e^{-\beta}\right)^4}\,,\\
f^{\phantom{g}}_{\text{C,non-min}|0}(\b) &=\frac{16\,e^{-2\,\b}}{\left(1-e^{-\beta}\right)^4}\,.
\end{split}
\end{equation}
The above functions are free from any singularity
in the positive $\b$ axis except for a pole at $\b=0$\,.
Series expanding the above, 
we find the coefficients,
\begin{equation}
\gamma^{\phantom{g}}_{\text{C,non-min}|2} = \frac{14}{15}\,,\qquad 
\gamma^{\phantom{g}}_{\text{C,non-min}|1}= \frac{4}{9}\,,\qquad 
\gamma^{\phantom{g}}_{\text{C,non-min}|0}= 0\,,
\end{equation}
which may be summed to finally give
\begin{equation}\label{vacnonmin}
\Gamma^{\rm\sst(1)\,\text{ren}}_{\text{C,non-min}} = \frac{62}{45}\,\log R\,.
\end{equation}
This matches to the result obtained in \cite{Beccaria:2014xda} by 
summing individual vacuum energies 
with damping factor $e^{-\e\,(s+\frac12)}$\,.
The result is precisely the ``vacuum energy'' of
a complex Maxwell field\footnote{The AdS$_5$ vacuum energy of the boundary spin-one
can be formally computed by interpreting the character of the short representation $\cS^1$ as the partition function for a field in AdS$_5$,
even though this primary does not represent a propagating degree of freedom in the bulk.
 A similar phenomenon was observed for the boundary scalar in \cite{Bae:2016rgm} where this bulk computation yielded an answer which reproduced the $a$ anomaly for the boundary conformal scalar. Using the partition function $\chi_{\cS_1}$, we obtain the expressions
\be
f_{\cS_1|2}(\b) =
f_{\cS_1|1}(\b) = \frac{e^{-2 \beta }-16\,e^{-\beta }+15}{24}\,,
\qquad
f_{\cS_1|0}(\b) =1.
\ee
These in turn may be expanded in powers of small $\beta$ to obtain
\begin{equation}
\gamma_{\cS^1|2} = \frac{7}{15}\,,\qquad \gamma_{\cS^1|1}= \frac{2}{9},\qquad \gamma_{\cS^1|0}= 0,
\end{equation}
which may be summed to finally obtain
\begin{equation}
\Gamma^{\sst(1)\,\text{ren}}_{\cS^1} = \frac{31}{45}\,\log R\,.
\label{vacsingl}
\end{equation}
This is consistent
with that boundary one-loop result \eqref{fuvdiv} 
upon using the correspondence between UV and IR in AdS/CFT dualities.} ---
which is twice of that of a real Maxwell field 
(that is, the spin-one doubleton $\cS_1$),
\be
    \Gamma^{\rm\sst(1)\,\text{ren}}_{\cS_1} = \frac{31}{45}\,\log R\,.
\ee

Now we turn to the interpretation of this result, as given in \cite{Beccaria:2014xda}, following \cite{Giombi:2013fka,Giombi:2014iua}. Firstly since we have a CFT with $N$ complex vectors, from equation \eqref{fuvdiv} we find that
\begin{equation}
F_{\sst U(N)\,\rm Maxwell} = 2\,N\, {31\over 45}\,\log\Lambda_{\rm\sst CFT}.
\end{equation}
Correspondingly, on the AdS side we have
\begin{equation}
\begin{split}
\Gamma_{\text{C,non-min}} &= \frac1{g}\,S_{\text{C,non-min}}+\Gamma^{\sst (1)\,\text{ren}}_{\text{C,non-min}}+\cO(g)\\
&=\frac1g\,S_{\text{C,non-min}} + {62\over 45}\,\log R +\mathcal{O}(g),
\end{split}
\end{equation}
where $S_{\text{C,non-min}}=\G^{\sst (0)}_{\text{C,non-min}}$ 
is the on-shell classical action with trivial boundary condition.
Using the correspondence between UV and IR divergences in the CFT and AdS theories respectively, we get
\begin{equation}
{62\over 45}\,N\,\log R = \frac1g\,S_{\text{C,non-min}} + {62\over 45}\,\log R\,,
\end{equation}
which suggests the identifications
\begin{equation}\label{typecnonmin}
    \frac1g= N-1\,,\qquad S_{\text{C,non-min}} = {62\over 45}\,\log R\,.
\end{equation}
Hence, the result \eqref{vacnonmin} suggests 
that the bulk inverse coupling constant should be shifted
from $N$ to $N-1$\,.

\subsubsection*{O(N)/Minimal Model}

We next turn to the minimal theory, whose character is given by
\be 
	\chi^{\phantom{g}}_{\rm C,min}(\b,\a_+,\a_-)=\frac{\chi^{\phantom{g}}_{\cS_1}(\b,\a_+,\a_-)^2
	+\chi^{\phantom{g}}_{\cS_1}(2\,\b,2\,\a_+,2\,\a_-)}2\,.
\ee
The contribution of the first term has already been evaluated
when we analyze the non-minimal theory in the previous section (see \eqref{vacnonmin}). 
We will therefore concentrate on the second term,
$\chi^{\phantom{g}}_{h}(\b,\a_+,\a_-)=\chi^{\phantom{g}}_{\cS_1}(2\,\b,2\,\a_+,2\,\a_-)$
up to 1/2 factor.
For this again, the functions $f_{h|n}$ may be computed
and then series expanded to obtain the quantities $\gamma_{h|n}$. 
We find
\be
\gamma_{h|2}= \frac{253}{480}\,,
\qquad 
\gamma_{h|1} = \frac{101}{288}\,,
\qquad
\gamma_{h|0} = {1\over 2}\,,
\ee
hence
\begin{equation}
\gamma_{h|2} + \gamma_{h|1} + \gamma_{h|0} = {62\over 45}\,.
\end{equation}
We thus finally obtain the one-loop vacuum energy of the minimal theory to be given by
\begin{equation}\label{vacmin}
\Gamma^{\sst(1)\,\text{ren}}_{\text{C,min}} = {1\over 2}\left(\frac{62}{45}+\frac{62}{45}\right)\log R= \frac{62}{45}\,\log R\,,
\end{equation}
which is same as that of non-minimal model \eqref{vacnonmin}.
This result may be interpreted in a similar manner to the non-minimal case above \cite{Beccaria:2014xda}. We again use \eqref{fuvdiv} to obtain
\begin{equation}
F_{\sst O(N)\,\rm Maxwell} = N\,{31\over 45}\,\log\Lambda_{\sst\rm CFT}.
\end{equation}
Correspondingly, on the AdS side we have
\begin{equation}
\begin{split}
\Gamma_{\text{C,min}} &= \frac1g\,S_{\text{C,min}}+\Gamma^{\sst (1)\,\text{ren}}_{\text{C,min}}+\cO(g)\\
&=\frac1g\,S_{\text{C,min}} + {62\over 45}\,\log R +\cO(g)\,,
\end{split}
\end{equation}
where $S_{\text{C,min}}=\G^{\sst (0)}_{\text{C,min}}$ 
is the on-shell classical action with trivial boundary condition.
We are thus led to identify
\begin{equation}
{31\over 45}\,N\,\log R = \frac1g\,S_{\text{C,min}} + {62\over 45}\,\log R\,,
\end{equation}
which suggests the identifications
\begin{equation}\label{typecmin}
\frac1{g}= N-2\,,\qquad S_{\text{C,min}}= {31\over 45}\,\log R\,.
\end{equation}
This suggests the bulk inverse coupling constant
should be shifted from $N$ to $N-2$\,.

\section{Holography for Free Yang-Mills}\label{freeym}

We now study the main target of the current work,
the holography of free Yang-Mills.
We begin with the analysis of field/operator content
of this AdS/CFT duality, in particular providing detailed discussions
on the first massive higher-spin multiplet, 
that we refer as to the second Regge trajectory.
We then  elaborate further the bulk theory based on 
its higher-spin symmetry and representations.
Consideration of the CFT correlators give additional hints about this theory.
Finally, we turn to the computation of the vacuum energy of this theory.

\subsection{Free $SU(N)$ Yang-Mills and its Single-Trace Operators} 
\label{sec: single trace}

The model we consider is the free limit of $SU(N)$ Yang-Mills.
In canonical normalization, the action is given by
\be
    S_{\rm CFT}=\frac14\int d^4x\,\tr\left[\bm F_{ab}\,\bm F^{ab}\right],
\ee 
where $\bm F_{ab}=2\,\partial_{[a} \bm A_{b]}$
and the gauge field $\bm A_a$ takes value in the adjoint representation
of $SU(N)$\,.
This literally free theory can be
also considered as the zero 't-Hooft coupling limit
in the large $N$ expansion of ordinary $SU(N)$ Yang-Mills. 
The single-trace operators are constructed from $\bm F_{ab}$
hence can be decomposed into the number of $\bm F_{ab}$ involved.
For instance, 
the single-trace operators at the order $k$ take the schematic form of
\be
    \tr\left[\partial^{n_1} \bm F\,\partial ^{n_2} \bm F\,\cdots \,\partial^{n_k} \bm F\right].
\ee
The spectrum of these operators can be obtained
by decomposing tensor products of $\cS_1$ into $so(2,4)$ UIRs.
In this adjoint case, the power of products can be increased to $N$
and in order to take into account of the cyclic symmetry of the trace,
the cyclic tensor products of $\cS_1$ have to be considered,
\be \label{s1tensorproduct}
    \cH_{k}=\Big[\overbrace{\cS_1\otimes \cdots \otimes \cS_1}^{k}\Big]_{\rm cyc}.
\ee
The cyclic tensor product can be realized combinatorically by making use 
of the P\'olya enumeration theorem, which defines the cyclic index
 \cite{Polyakov:2001af,Bianchi:2003wx,Beisert:2003te,Spradlin:2004pp}.
The latter is nothing but the character of $\cH_k$
and it is given by
\be
    \chi_{{\rm cyc}^k}(g)
    =\frac1k\,\sum_{n|k}\,\varphi(n)\left[\chi_{\cS_1}(g^n)\right]^{\frac{k}{n}}\,,
    \label{cyc char}
\ee
where $n|k$ indicates the positive integers $n$ which divide $k$,
and $\varphi(n)$ is the Euler totient function, which counts
the number of relative prime of $n$ in $\{1,2,\ldots,n\}$\,.

As we have noted previously, the spectrum of 
single-trace operators 
in adjoint model CFTs is immensely larger than that of vector models, and indeed one important virtue in our method of computing the one-loop vacuum energy is that a precise determination of the spectrum is not required. 
Nonetheless, it would be of interest to explicate the spectrum, at least for some small powers of $k$ appearing in the cyclic tensor product \eqref{s1tensorproduct}. 
Firstly we note that at the order $k=2$, the cyclic product is same as the symmetric one, hence the single-trace operators at this order are the same as those of free Maxwell $O(N)$ model. The spectrum obtained from these operators has already been enumerated in \eqref{cyclicj1spec}. Hence we directly proceed to the case of $k=3$.

\subsubsection*{Order Three Single-Trace Operators}

Let us consider the spectrum of single-trace operators
of free Yang-Mills involving three $\bm F_{ab}$\,.
These also define the field content of the AdS theory
in the second Regge trajectory.
The relevant character is the $k=3$ case of \eqref{cyc char}
and given by
\begin{equation}
\chi_{{\rm cyc}^3}(\b,\a_+,\a_-)
={\chi_{\cS_1}\!\left(\beta,\alpha_+,\alpha_-\right)^3 +2\,\chi_{\cS_1}\!\left(3\,\beta,3\,\alpha_+,3\,\alpha_-\right)\over 3}.
\label{2RT char}
\end{equation}
To get the operator/field content, we decompose
the above character into that of $so(2,4)$ UIRs. 
It may be possible, though we do not do so here, to carry out an explicit decomposition of the representation by means of an oscillator construction, as carried out in \cite{Gunaydin1998,Bae:2016rgm}.
Instead, we use here the method developed in \cite{Newton:2008au} and reviewed in 
Section \ref{sec: NS}.
The formula \eqref{genfunc} 
gives the generating function of the multiplicities
of the operator/field content as
\be
    G_{{\rm cyc}^3}\!\left(q,\frac{x_1}q,x_2\right)
    =\sum_{\t=3}^\infty q^\t\,G_{{\rm cyc}^3}^\t(x_1,x_2)\,,
\ee
where $x_1^2=x_+\,x_-$ and $x_2^2=x_+/x_-$\,,
hence the power of $x_1$ and $x_2$ correspond
to $\ell_1$ and $\ell_2$ of the two-row Young diagram of $so(4)$\,.
Here, we expand the generating function $G_{{\rm cyc}^3}$
in terms of twist $\t=\D-\ell_1$
since it organizes the spectrum better
than the conformal weight (or lowest energy) $\D$\,.
The multiplicity generating function of given twist $\t$
can be again expanded as a series in $x_2$,
\be
    G_{{\rm cyc}^3}^\t(x_1,x_2)=
    \sum_{\ell_2=0}^{\t} {x_2}^{\ell_2}\,
    G_{{\rm cyc}^3}^{\t,\ell_2}(x_1)\,.
\ee
Here, we consider only positive powers of $x_2$ as 
the theory is parity-invariant: the coefficients of negative powers
are identical to those of positive ones.
Now, let us provide explicit examples for a few lower twists.
In the second Regge trajectory, the lowest twist is three, hence
only contains long representation, that is, massive fields in AdS$_5$.

\paragraph{Twist Three}

At twist three, we get
\be
    G_{{\rm cyc}^3}^3(x,y)
    =y\,G_{{\rm cyc}^3}^{3,1}(x)
    +y^3\,G_{{\rm cyc}^3}^{3,3}(x)\,,
\ee
hence, there are two types of fields:
\be
\parbox{70pt}{
	\begin{tikzpicture}
	\draw (0,0) rectangle (2.4,0.4);
	\node at (1.2,0.2){$\ell_1$};
	\draw (0,0) -- (0,-0.4) -- (0.4,-0.4) -- (0.4,0);
	\end{tikzpicture}}\,,
	\qquad 
\parbox{70pt}{
	\begin{tikzpicture}
	\draw (0,0) rectangle (2.4,0.4);
	\node at (1.2,0.2){$\ell_1$};
	\draw (0,0) -- (0,-0.4) -- (1.2,-0.4) -- (1.2,0);
	\draw (0.4,0) -- (0.4,-0.4);
	\draw (0.8,0) -- (0.8,-0.4);
	\end{tikzpicture}}\,.
\ee
The first type of fields with $\ell_2=1$ have
the multiplicity generating function
\be
     G_{{\rm cyc}^3}^{3,1}(x)
     =\frac{x^3}{(1-x)^2}
     =\sum_{\ell_1=3}^\infty\,(\ell_1-2)\,x^{\ell_1}\,,
\ee 
or in other words, the UIR decomposition contains
\be
    \bigoplus_{\ell_1=3}^\infty\,(\ell_1-2)\,
    \cD(3+\ell_1,(\ell_1,1)_{\rm\sst PI})\,.
\ee
The multiplicities of the second type fields with $\ell_2=3$
are generated by
\ba
     &&G_{{\rm cyc}^3}^{3,3}(x)
     =\frac{x^3\,(1-x+x^2)}{(1-x)^2\,(1+x+x^2)}\nn
     && =x^3+x^5+2\,x^6+x^7+2\,x^8+3\,x^9+2\,x^{10}+3\,
     x^{11}+\cO(x^{12})\,.
\ea

\paragraph{Twist Four}

The fields of twist four are of particular interest
as they include fields that can serve as Goldstone modes for massless HS fields.
We remind the reader that all the massless fields in AdS$_5$
have the twist two and the Goldstone modes have the twist
greater than that of massless fields by two, hence four.
In this case, we get
\be
    G_{{\rm cyc}^3}^4(x,y)
    =G_{{\rm cyc}^3}^{4,0}(x)
    +y^2\,G_{{\rm cyc}^3}^{4,2}(x)
    +y^4\,G_{{\rm cyc}^3}^{4,4}(x)\,,
\ee
hence, there are four types of fields:
\be
\parbox{70pt}{
	\begin{tikzpicture}
	\draw (0,0) rectangle (2.4,0.4);
	\node at (1.2,0.2){$\ell_1$};
	\end{tikzpicture}}\,,
	\qquad 
\parbox{70pt}{
	\begin{tikzpicture}
	\draw (0,0) rectangle (2.4,0.4);
	\node at (1.2,0.2){$\ell_1$};
	\draw (0,0) -- (0,-0.4) -- (0.8,-0.4) -- (0.8,0);
	\draw (0.4,0) -- (0.4,-0.4);
	\end{tikzpicture}}\,,
	\qquad 
\parbox{70pt}{
	\begin{tikzpicture}
	\draw (0,0) rectangle (2.4,0.4);
	\node at (1.2,0.2){$\ell_1$};
	\draw (0,0) -- (0,-0.4) -- (1.6,-0.4) -- (1.6,0);
	\draw (0.4,0) -- (0.4,-0.4);
	\draw (0.8,0) -- (0.8,-0.4);
	\draw (1.2,0) -- (1.2,-0.4);
	\end{tikzpicture}}\,.
\ee
First, the multiplicities of symmetric fields are generated by
\ba
     && G_{{\rm cyc}^3}^{4,0}(x)
     =\frac{2\,x^2}{(1-x)^2}
     =\sum_{\ell_1=2}\,2\,(\ell_1-1)\,x^{\ell_1}\nn 
     &&=2\,x^2+4\,x^3+
     6\,x^4+8\,x^5+10\,x^6+12\,x^7+14\,x^8+16\,x^{9}+\cO(x^{10})\,.
\ea 
These include all the Goldstone modes $\cD(s+3,(s-1,0))$
for symmetric massless HS fields $\cD(s+2,(s,0))$
except for the $s=1$ and 2, namely the Goldstone modes for 
a $U(1)$ gauge field and graviton
--- higher Regge trajectory fields 
do not contain such modes either.
The absence of Goldstone modes for the U(1) gauge field and the graviton is interesting, and understandable as this suggests that when a coupling is weakly turned on for the CFT, higgsing the higher-spin symmetry, the graviton and U(1) gauge field do not get a mass term as there are no Goldstone bosons they can eat.

The second type of fields have the multiplicities
correspondingly to
\ba
     &&G_{{\rm cyc}^3}^{4,2}(x)
     =\frac{2\,x^3\,(1+x)}{(1-x)^2\,(1+x+x^2)}\nn
     && =2\,x^3+4\,x^4+4\, x^5+6\, x^6+8\, x^7+8\, x^8+
     10\, x^9+
     \cO(x^{10})\,.
     \label{G 4 2}
\ea
They include the Goldstone modes 
$\cD(s+3,(s-1,2)_{\rm\sst PI})$
for mixed-symmetry HS fields $\cD(s+2,(s,2)_{\rm\sst PI})$
except for $s=3$ case.
Field of this exceptional case does not appear in the first Regge trajectory, hence all mixed-symmetry massless fields in the spectrum
can acquire mass through a Higgs mechanism.

The last type of fields with $\ell_2=4$
have the multiplicity generating function,
\ba
     &&G_{{\rm cyc}^3}^{4,4}(x)
     =\frac{x^4\,(1-x+x^2)}{(1-x)^2\,(1+x+x^2)}\nn
     && =x^4+x^6+2\,x^7+x^8+2\,x^9+3\,x^{10}+2\,x^{11}
     +3\,
     x^{12}+\cO(x^{13})\,.
\ea
These fields can be also considered as Goldstone
modes $\cD(s+3,(s-1,4)_{\rm\sst PI})$ but
the corresponding massless fields are not present 
in the theory.

\paragraph{Higher Twist}

In principle, one can proceed to higher twist in
the same manner. For instance, the multiplicities of 
fields
having twist five and six are all encoded in the generating functions,
\be
    G_{{\rm cyc}^3}^{5}(x,y)
    =
    y\,\frac{2\,x\,(1+x)}{(1-x)^2\,(1+x+x^2)}
    +y^3\,\frac{x^3}{(1-x)^2}
    +y^5\,\frac{x^5\,(1-x+x^2)}{(1-x)^2\,(1+x+x^2)}\,,
\ee
and
\be
    G_{{\rm cyc}^3}^{6}(x,y)
    =
    \frac{2\,(1-x+x^2)}{(1-x)^2\,(1+x+x^2)}
    +y^2\,\frac{x^2}{(1-x)^2}
    +y^4\,\frac{x^4}{(1-x)^2}
    +y^6\,\frac{x^6\,(1-x+x^2)}{(1-x)^2\,(1+x+x^2)}\,.
\ee
For higher twists, it turns out that 
there is no limit of 
allowed twists in the second
Regge trajectory as opposed to
the case of scalar adjoint model in AdS$_4$ \cite{Bae:2016rgm}.
For a fixed twist $\t$\,,
at most four values of $\ell_2$
appear: \mt{\ell_2= \t, \t-2, \t-4, \t-6}\,.
Therefore, twist six is the maximum twist which contain symmetric HS fields ($\ell_2=0$).
Moreover, it is the only twist in the second Regge trajectories
which contains scalar fields:
there are two of them as
$G_{{\rm cyc}^3}^{6}(x,y)=2+\cO(x,y)$.

\subsection{AdS Dual of Free Yang-Mills}

With the above inputs, let us now turn to the putative Bulk theory
Dual to free Yang-Mills (BDYM).

\subsubsection*{Field Content}

The first hint about the theory is its field content. In the planar limit, this may be identified
with the single-trace operator spectrum of free Yang-Mills. As we have discussed in the previous section,
this spectrum can be arranged into different Regge trajectories (RT),
where the $n$-th Regge trajectory, which we denote as RT$_n$, contains fields dual to the operator involving $n+1$ 
power of the Yang-Mills curvature $\bm F_{ab}$\,. The fields in the first trajectory
coincide with those of the Type C theory
as the operator spectrum does so on the CFT side.
Let us remind the reader that this trajectory
contains infinitely massless fields
of symmetric type,
\be
    \varphi_{\mu_1\cdots \mu_s}\qquad [s=2,3,\ldots,\infty]\,,
    \label{sym}
\ee
as well as the mixed-symmetry fields,
\be
    \varphi_{\mu_1\cdots \mu_s,\nu_1\nu_2}\qquad 
    [s=4,6,\ldots,\infty]\,,
    \label{mixedsym f}
\ee
of the Young diagram type $(s,2)_{\rm\sst PI}$\,.
Remark that the fields \eqref{mixedsym f} are only for even $s\ge 4$\,. 
There are also four non-gauge fields in this trajectory:
two scalar fields (actually one scalar and one pseudo-scalar) $\varphi$ and $\varphi'$, 
one \emph{non-gauge} two-form $\varphi_{[\m\n]}$ and one
$(2,2)$ mixed-symmetry field $\varphi_{\m_1\m_2,\n_1\n_2}$\,.
On top of the first Regge trajectory fields, there are infinitely many higher trajectories, each of which contains infinitely many fields.
All these fields are massive and 
their `masses' increase as the trajectory number $n$ grows.
This is because the minimum \emph{twist} ---
that is, the minimum energy $\D$ minus the `spin' $\ell_1$ ---
of the $n$-th trajectory fields is $n+1$,
whereas all massless fields in AdS$_5$ have twist 2.
These massive fields make use of all allowed mixed-symmetry tensors.
Let us also mention that
the field content of BDYM
is  a subset of the type IIB string theory spectrum in AdS$_5\times$S$^5$ 
background 
since the free Yang-Mills is a subset of free $\cN=4$ Super Yang-Mills.

\subsubsection*{Cubic Interactions and HS Symmetry}

Let us now discuss about the interaction structure of this  theory,
starting from that of cubic order.
First of all, non-Abelian cubic interactions of 
massless gauge fields should be fixed by an underlying  HS algebra.
Actually, such HS algebra
is a bosonic subalgebra of 
the maximal symmetry of free $\cN=4$ theory
\cite{Sezgin:2001yf,Sezgin:2001zs,Vasiliev:2001wa,Vasiliev:2001zy, 
 Beisert:2004di,Bianchi:2005ze}.
It consists of the generators 
\be
\parbox{70pt}{
	\begin{tikzpicture}
	\draw (0,0) rectangle (2.4,0.8);
	\draw (0,0.4) -- (2.4,0.4);
	\node at (1.2,0.6){$s-1$};
	\node at (1.2,0.2){$s-1$};
	\end{tikzpicture}}\,,
    \qquad
\parbox{70pt}{
	\begin{tikzpicture}
	\draw (0,0) rectangle (2.4,0.8);
	\draw (0,0.4) -- (2.4,0.4);
	\node at (1.2,0.6){$s-1$};
	\node at (1.2,0.2){$s-1$};
	\draw (0,0) -- (0,-0.4) -- (0.8,-0.4) -- (0.8,0);
	\draw (0.4,0) -- (0.4,-0.4);
	\end{tikzpicture}}\,,
\ee
corresponding to the Killing tensors
of symmetric and mixed-symmetry fields
$\varphi_{\m_1\cdots \m_s}$
and $\varphi_{\m_1\cdots \m_s,\n_1\n_2}$,
respectively.\footnote{See \cite{Alkalaev:2003qv}
for some generality about the mixed-symmetry fields and their Killing tensors.}
Its subalgebra consisting of two-row Young diagram
generators has been studied in the context 
of the conformal HS theory in \cite{Fradkin:1989yd}.
Oscillator realization of generic doubleton representations 
have been studied in \cite{Gunaydin1989,Sezgin:2001zs}.
More recent discussions 
can be found in \cite{Bekaert:2009fg,
Boulanger:2011se,Boulanger:2013zza,Govil:2013uta,Joung:2014qya}.

As the spin-one doubleton plays the role of 
fundamental representation of this HS algebra,
any of its tensor products  forms
a multiplet under the same algebra. 
This means that the quadratic action of these fields 
belonging to a given, say $n$-th, Regge trajectory should admit such HS symmetry.
This symmetry acts on these fields in a way mixing different components and fields, somewhat similarly to supersymmetry.
This rigid HS symmetry can be gauged by coupling the corresponding  currents, made by the $n$-th trajectory fields, to symmetric HS gauge fields.
In addition, this HS symmetry would restrict even the rest of 
cubic interactions, involving three curvatures, sometime referred as
`Born-Infeld-type'. Therefore, in the end, 
all the cubic interactions will be fully determined by HS symmetry. 
This is not surprising because the cubic interactions of the AdS theory
can be determined by the three-point functions in CFT,
which in turn are fixed uniquely by the HS symmetry \cite{Alba:2013yda}.

In fact, by considering possible CFT three-point functions, one
can extract additional information about the AdS cubic interactions. 
Denoting CFT operators involving $n$ powers of $\bm F_{ab}$
as $\cO^{{\rm RT}_n}$\,, one can 
conclude that the connected part of the
three-point function,
\be
    \la \cO^{{\rm RT}_{n_1}}\,\cO^{{\rm RT}_{n_2}}\,\cO^{{\rm RT}_{n_3}}\ra
    \qquad [n_1\le n_2\le n_3]\,,
\ee 
vanishes if
\begin{equation}
     n_3-n_1-n_2>0\quad {\rm or \quad odd}\,.
    \label{cond abs}
\end{equation}
This is simply because in a free CFT, all correlation functions
are entirely fixed by Wick contractions.
This directly leads us to the following two conclusions about the cubic interactions of 
the AdS theory. Firstly, cubic interactions are absent for the fields belongings to the $(n_i-1)$-th
RT, which we denote by RT$_{n_i}$, satisfying \eqref{cond abs}.
In particular, the massless-massless-massive interactions, that is 
RT$_2 -\!$ RT$_2 -\!$ RT$_{n}$ interactions,
exist only for $n=4$\,. This implies that the massless gauge fields can
source only the fields of the third  Regge
trajectories. Further, two fields in any RT$_n$ can source a single field in RT$_2$. That is, RT$_2-\!$ RT$_n -\!$ RT$_{n}$ is an allowed interaction in the bulk for any $n$\,.
Secondly, all fields with odd $n$ can be truncated consistently at least at the cubic level.
This in fact makes sense from the CFT point of view: by considering CFT fields taking value in a non-square matrix space, we would restrict only to the
operators having even $n$\,.

\subsubsection*{Quartic Interaction and Locality}

Quartic interaction is the pandora's box,
containing many crucial subtleties of HS field theory. In the unfolded formulation of Vasiliev theory,
the quartic information is not manifest, but
once presented in a conventional metric-like form, 
even the scalar field turns out to involve non-local
quartic coupling,
as recently shown in \cite{Bekaert:2014cea,Bekaert:2015tva} using AdS/CFT duality
(see \cite{Sleight:2016dba} for the cubic order analysis). 
In the holographic correspondence, the quartic interactions enter in the determination
of four-point Witten diagrams, which are in turn directly related to the CFT four-point function.
Hence, quartic interactions can be identified from CFT four-point function
by extracting the exchange diagram contributions.
Returning to BDYM,
its quartic interactions of the first Regge trajectory fields
can be obtained from the four-point function
of the operators quadratic in $\bm F_{ab}$
by removing
the contribution of exchange diagrams
where the exchanged fields belong to either
the first or the third RT:
\be
	\parbox{80pt}{
	\begin{tikzpicture}
	\draw [blue, semithick] (0,0) circle [radius=1.2];
    \draw [semithick] (45:1.2) -- (45:-1.2);
    \draw [semithick] (-45:1.2) -- (-45:-1.2);
    \node at (45:1.5) {\scriptsize RT$_2$};
    \node at (-45:1.5) {\scriptsize RT$_2$};
    \node at (45:-1.5) {\scriptsize RT$_2$};
    \node at (-45:-1.5) {\scriptsize RT$_2$};
	\end{tikzpicture}}
	=\ 
	(4\ {\rm pt})
	\ -\ 
	\parbox{80pt}{
	\begin{tikzpicture}
	\draw [blue, semithick] (0,0) circle [radius=1.2];
    \draw [semithick] (45:1.2) -- (0.4,0) -- (-45:1.2);
    \draw [semithick] (0.4,0) -- (-0.4,0);
    \draw [semithick] (45:-1.2) -- (-0.4,0) -- (-45:-1.2);
    \node at (45:1.5) {\scriptsize RT$_2$};
    \node at (-45:1.5) {\scriptsize RT$_2$};
    \node at (45:-1.5) {\scriptsize RT$_2$};
    \node at (-45:-1.5) {\scriptsize RT$_2$};
    \node at (0,-0.25) {\scriptsize RT$_2$};
	\end{tikzpicture}}
	 \,-\ 
	\parbox{80pt}{
	\begin{tikzpicture}
	\draw [blue, semithick] (0,0) circle [radius=1.2];
    \draw [semithick] (45:1.2) -- (0.4,0) -- (-45:1.2);
    \draw [semithick] (0.4,0) -- (-0.4,0);
    \draw [semithick] (45:-1.2) -- (-0.4,0) -- (-45:-1.2);
    \node at (45:1.5) {\scriptsize RT$_2$};
    \node at (-45:1.5) {\scriptsize RT$_2$};
    \node at (45:-1.5) {\scriptsize RT$_2$};
    \node at (-45:-1.5) {\scriptsize RT$_2$};
    \node at (0,-0.25) {\scriptsize RT$_4$};
	\end{tikzpicture}}.
\ee
Now the second diagram on the right hand side does not exist in the Type C theory, as only fields in RT$_2$ appear there. This shows that
the quartic coupling of massless fields in this stringy AdS 
theory of BDYM would generically differ from that of Type C theory. 
The same conclusion can be drawn for the
AdS duals of free scalar adjoint and vector models.
It will be interesting to
compare the quartic interactions of these theories,
that is,
one dual to vector model and the other dual to adjoint model CFT.

\subsubsection*{HS Higgs Mechanism}

Probably one of the most interesting and challenging
issues in the duality of free Yang-Mills 
is the question of extending the duality to interacting Yang-Mills. This putative correspondence
has been extensively studied, but the focus 
was mostly on the understanding of the
strongly coupled boundary theory.
Here, we take a more modest point of view:
we assume to extend the free Yang-Mills duality
to weakly interacting one, hence in high energy.
From AdS point of view, the background geometry
is nearly AdS in the asymptotic region but drastically modified in deep interior region. 
This modification of the background should be caused by
a kind of Higgs mechanism, where HS symmetry is broken down 
to boundary Poincar\'e symmetry $iso(1,3)$.
The breaking of HS symmetry makes all massless HS fields
acquire masses. More precisely, their UIR
$\cD(s+2,(s,0))$ and $\cD(s+2,(s,2)_{\rm\sst PI})$
should be combined with the Goldstone modes,
$\cD(s+3,(s-1,0))$ and $\cD(s+3,(s-1,2)_{\rm\sst PI})$\,,  respectively
\cite{Bianchi:2005ze}.
These modes are related to the gauge mode
$\cD(3,(1,0))$ of the doubleton $\cD(2,(1,1)_{\rm\sst PI})$\,,
which can be interpreted as the right side
of interacting Yang-Mills equations:
\be
    \partial^{a}\,\bm F_{ab}
    =g\,\Big([\bm A^a,\bm F_{ab}]
    +\partial^a\,[\bm A_a\,,\bm A_b]\Big)
    +\mathcal{O}(g^2)\,,
\ee
where $\bm F_{ab}=2\,\partial_{[a}\bm A_{b]}$ 
is the curvature of free theory.
These contributions also give rise to the right hand
side of the current conservation condition,
\be
    \partial^{a_1}\,J_{a_1\cdots a_s}
    =g\,K_{a_1\cdots a_{s-1}}+\mathcal{O}(g^2)\,,
    \qquad 
    \partial^{a_1}\,J_{a_1\cdots a_s,b_1b_2}
    =g\,K_{a_1\cdots a_{s-1},b_1b_2}+\mathcal{O}(g^2)\,,
\ee
where $K_{a_1\cdots a_{s-1}}$ and $K_{a_1\cdots a_{s-1},b_1b_2}$
are the operators made by three $\bm F_{ab}$\,.
They are dual to the HS Goldstone fields belonging to the second RT,
which contains also other massive fields.
Let us denote the Goldstone fields and the others by
GS and RT$_3'$, respectively.
They altogether form a massive multiplet of HS symmetry.
The mass generation mechanism itself should be
controlled in particular by the cubic interaction 
RT$_2-$RT$_3-$RT$_3$. 
The second Regge trajectory RT$_3$ splits into the Goldstone fields 
GS and the rest of the fields 
RT$_3'$. When the latter takes
a non-trivial background value proportional to $g$, the cubic coupling 
generates the kinetic term mixing between RT$_2$ and GS
which realize a Stueckelberg formulation of massive HS:
\be
    \parbox{80pt}{
	\begin{tikzpicture}
    \draw [semithick] (90:1) -- (0,0);
    \draw [semithick] (210:1) -- (0,0);
    \draw [semithick] (-30:1) -- (0,0);
    \node at (90:1.2) {\scriptsize RT$_2$};
    \node at (-30:1.3) {\scriptsize RT$_3$};
    \node at (210:1.2) {\scriptsize RT$_3$};
	\end{tikzpicture}}
    \Rightarrow 
    \parbox{80pt}{
	\begin{tikzpicture}
    \draw [semithick] (90:1) -- (0,0);
    \draw [semithick] (210:1) -- (0,0);
    \draw [dotted] (-30:1) -- (0,0);
    \node at (90:1.2) {\scriptsize RT$_2$};
    \node at (-30:1.3) {\scriptsize RT$_3'$};
    \node at (210:1.3) {\scriptsize GS};
	\end{tikzpicture}}\,,
\ee
and the massless fields in RT$_2$ acquire
masses proportional to $g^2$\,.
Moreover the background value  of 
RT$_3'$ would source
the graviton such that the gravitational background is also
deformed to a non-AdS geometry:
\be
    \parbox{80pt}{
	\begin{tikzpicture}
    \draw [semithick] (90:1) -- (0,0);
    \draw [dotted] (210:1) -- (0,0);
    \draw [dotted] (-30:1) -- (0,0);
    \node at (90:1.2) {$\st g_{\m\n}$};
    \node at (-30:1.3) {\scriptsize RT$_3'$};
    \node at (210:1.3) {\scriptsize RT$_3'$};
	\end{tikzpicture}}\,,
\ee
whereas in the case of String theory dual to $\cN=4$ theory,
the source term should add up to vanish.
See \cite{Bianchi:2005ze} for the discussions of Higgs mechanism there.
Let us also note that, in the latter case, 
the vacuum expectation value should not 
break the AdS covariance, and the only fields which may fulfill this condition
are scalar and symmetric rank-two fields,
when formulated in terms of doubly traceless fields.
As a final observation, 
the only massless-massless-massive interactions
are  RT$_2-$RT$_2-$RT$_{4}$.
Therefore, 
it is possible to turn on background values for fields in RT$_{n\geq 5}$ without inducing 
additional mass-like terms for RT$_2$\,.
Also, since vertices of the form RT$_{2}-$RT$_{n_2}-$RT$_{n_3}$ are generically allowed, turning on background values for fields in massive RTs would generically source the graviton, deforming the background to a non-AdS geometry.
Of course, this perturbative picture should be 
appended by all higher order corrections
to realize fully non-perturbative Higgs mechanism.

\subsection{One-Loop Vacuum Energy in AdS$_5$} 
\label{freeymtest}

We now compute the one-loop vacuum energy 
of Bulk Dual theory of free $SU(N)$ Yang-Mills 
(BDYM) around AdS$_5$ with $S^4$ boundary. 
Compared to the spectrum of the Type-C HS theories,
this  HS theory has a vastly extended field content as we have discussed before. 
The field content is encoded in the character (or generalized partition function) of the CFT, which can be in turn determined as in \eqref{cyc char}.
Hence, it is possible to apply the CIRZ method
to compute the zeta function of  
BDYM, and thus the one-loop vacuum energy.

\subsubsection{Zeta Functions for 
First Few Regge Trajectories}\label{higherregge}

In this section we shall compute
the one-loop vacuum energies of the fields in the first few Regge trajectories of BDYM. 
We begin with 
rather explicit demonstration of the results
for the first three Regge trajectories
then  display the higher order quantities graphically. 
The CIRZ method requires explicit form
of character for each
Regge trajectories,
and they are given in \eqref{cyc char}. 
For completeness, we start with the order two contribution.

\subsubsection*{Order Two}

This is the same as the minimal Type-C computation, which has already been carried out. The answer obtained is
\begin{equation}
\Gamma^{\sst (1)\,\text{ren}}_{\text{cyc}^2} = {62\over 45}\,\log R\,,
\end{equation}
which is twice of $\Gamma^{\sst (1)\,\text{ren}}_{\cS_1}$,
the `vacuum energy' of the boundary spin-one field.

\subsubsection*{Order Three}

The order three character is given by
\eqref{2RT char}.
From it, we may compute $f_{\text{cyc}^3|n}\!\left(\beta\right)$ and its corresponding series expansion about $\beta=0$ to obtain
\begin{equation}
\gamma_{\text{cyc}^3|2}=\frac{55141}{51975}\,,
\qquad \gamma_{\text{cyc}^3|1} = \frac{193541}{299376}\,,\qquad \gamma_{\text{cyc}^3|0} = \frac{665702}{1403325}\,.
\end{equation}
We may sum these to obtain
\begin{equation}
\Gamma^{\sst (1)\,\text{ren}}_{\text{cyc}^3} = \frac{4453429}{2041200}\,\log R \simeq 2.18177\,\log R.
\end{equation}
As before, we compare this to the boundary spin-one result
to obtain
\begin{equation}
{\Gamma^{\sst (1)\,\text{ren}}_{\text{cyc}^3}\over \Gamma^{\sst (1)\,\text{ren}}_{\cS_1}}=\frac{143659}{45360}\simeq 3.16709\,.
\label{order 3}
\end{equation}

\subsubsection*{Order Four}

The order four character is given by
\begin{equation}
\chi_{\text{cyc}^4}\!\left(\beta,\alpha_+,\alpha_-\right) = {\chi_{\cS_1}\!\left(\beta,\alpha_+,\alpha_-\right)^4+ \chi_{\cS_1}\!\left(2\,\beta,2\,\alpha_+,2\,\alpha_-\right)^2 +2\,\chi_{\cS_1}\!\left(4\,\beta,4\,\alpha_+,4\,\alpha_-\right)
\over 4}.
\end{equation}
In that case we find 
\begin{equation}
\gamma_{\text{cyc}^4|2}=\frac{42641101}{26611200}\,,
\qquad \gamma_{\text{cyc}^4|1} = \frac{9204955}{9580032}\,,\quad \gamma_{\text{cyc}^4|0} = \frac{198713}{369600}\,,
\end{equation}
and hence
\begin{equation}
\Gamma^{\sst (1)\,\text{ren}}_{\text{cyc}^4} = {2109829\over 680400}\,\log R \simeq 3.10087\,\log R\,.
\end{equation}
Comparing this to the `vacuum energy'
of the boundary spin-one, we get
\begin{equation}
{\Gamma^{\sst (1)\,\text{ren}}_{\text{cyc}^4}\over \Gamma^{\sst (1)\,\text{ren}}_{\cS_1}}=\frac{68059}{15120}\simeq 4.50126\,.
\label{order 4}
\end{equation}

\subsubsection*{Higher Orders}

\begin{figure}[h]
\centering
\includegraphics[width=0.7\textwidth]{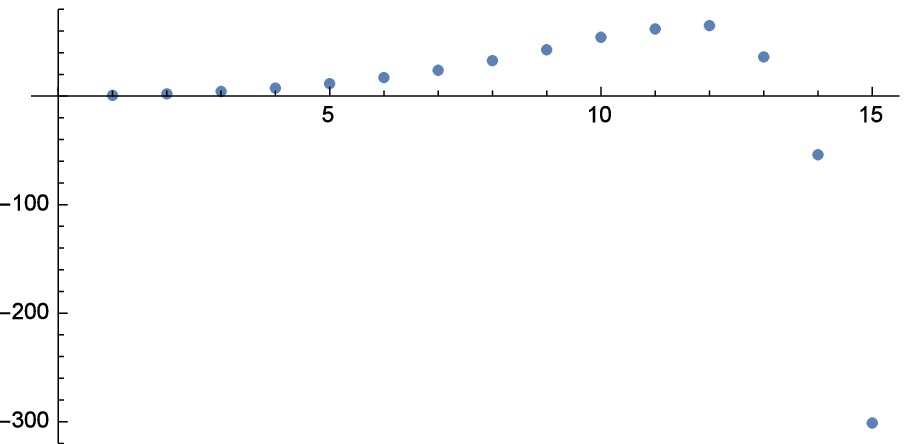}
\caption{Plot of $\sum_{n=2}^{N}\,
    \Gamma^{\sst (1)\,\text{ren}}_{\text{cyc}^n}$
upto $N=15$}
\label{Full vacenergy}
\end{figure}

We may push the computation of zeta functions 
to higher Regge trajectories. While we do not provide explicit answers here, we do exhibit the results graphically. 
\begin{figure}[h]
\centering
\includegraphics[width=0.7\textwidth]{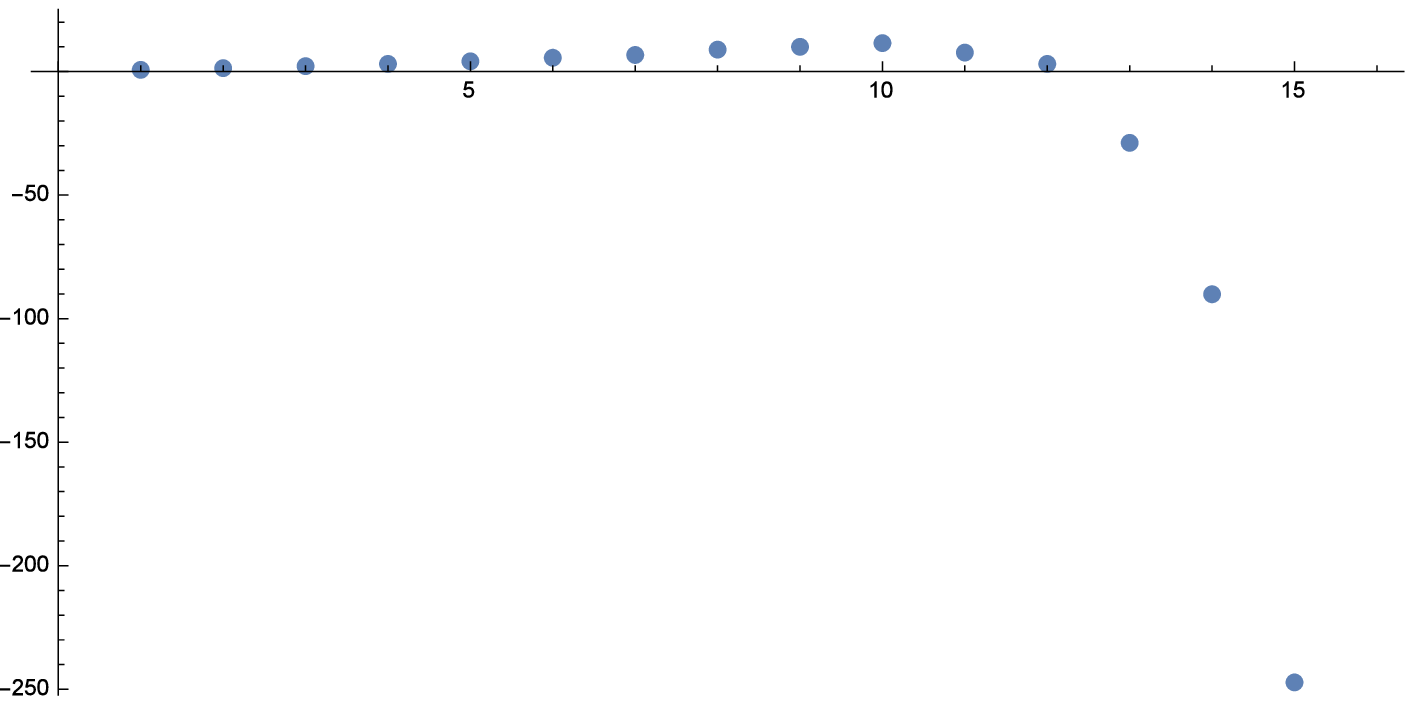}
\caption{Plot of
$\G^{\sst (1)\,\rm ren}_{{\rm cyc}^n}$
from $n=1$ to 15}
\label{vacenergy}
\end{figure}
Figure \ref{vacenergy} is  the plot of the one-loop vacuum energies
of the fields in the first 15th Regge trajectories.
It does not shows the curious pattern\footnote{
In the holography of free scalar adjoint model, the vacuum energies of the AdS fields in the first $32$ trajectories 
coincide, when normalized with Euler totient function, with that of boundary scalar within 1\% of error.} 
observed for the free scalar adjoint model \cite{Bae:2016rgm},
but there are other remarkable points:
first of all, the vacuum energies
seems to follow a smooth curve than a chaotic oscillation.
Moreover, the vacuum energy becomes negative
starting from $n=13$\,.
This sign flip of vacuum energy
is certainly remarkable.
Even though we do not have a clear physical interpretation for this phenomenon, 
let us elaborate a possible direction.
When $N$ is a finite number, the power of $\bm F_{ab}$ 
which can appear inside of a single trace is limited.
Only first $N-1$ Regge trajectories ($n=2,\ldots, N$) do not lose any of its field content.\footnote{The Regge trajectories with $N<n<N^2$ will have a partial loss of single-particle content, whereas higher trajectories lose the entire 
 content of single-particle fields.}
Taking only the contributions of the ``untouched'' trajectories,
we get the sum $\sum_{n=2}^{N}\,
    \Gamma^{\sst (1)\,\text{ren}}_{\text{cyc}^n}$\,,
    which are plotted in Figure\,\ref{Full vacenergy} for 
$N=2,\ldots,15$\,.
Now, we see that the sign flip happens
for particular value of $N$, \textit{viz.} $N=14$, which is an external parameter
of the theory. Would this show a sign of 
phase-transition in $N$?
Of course, one have to bear in mind that
we have not take into account the contributions from
the trajectories with $N<n<N^2$ and
the finite $N$ would 
nullify many nice features. The $1/N$ expansion becomes purely numerical, hence it is hard to assume that the lower order in the expansion can be considered seperately. Moreover, the assumed map between CFT operators and AdS fields would get a finite correction.

\begin{figure}[h]
\centering
\includegraphics[width=0.7\textwidth]{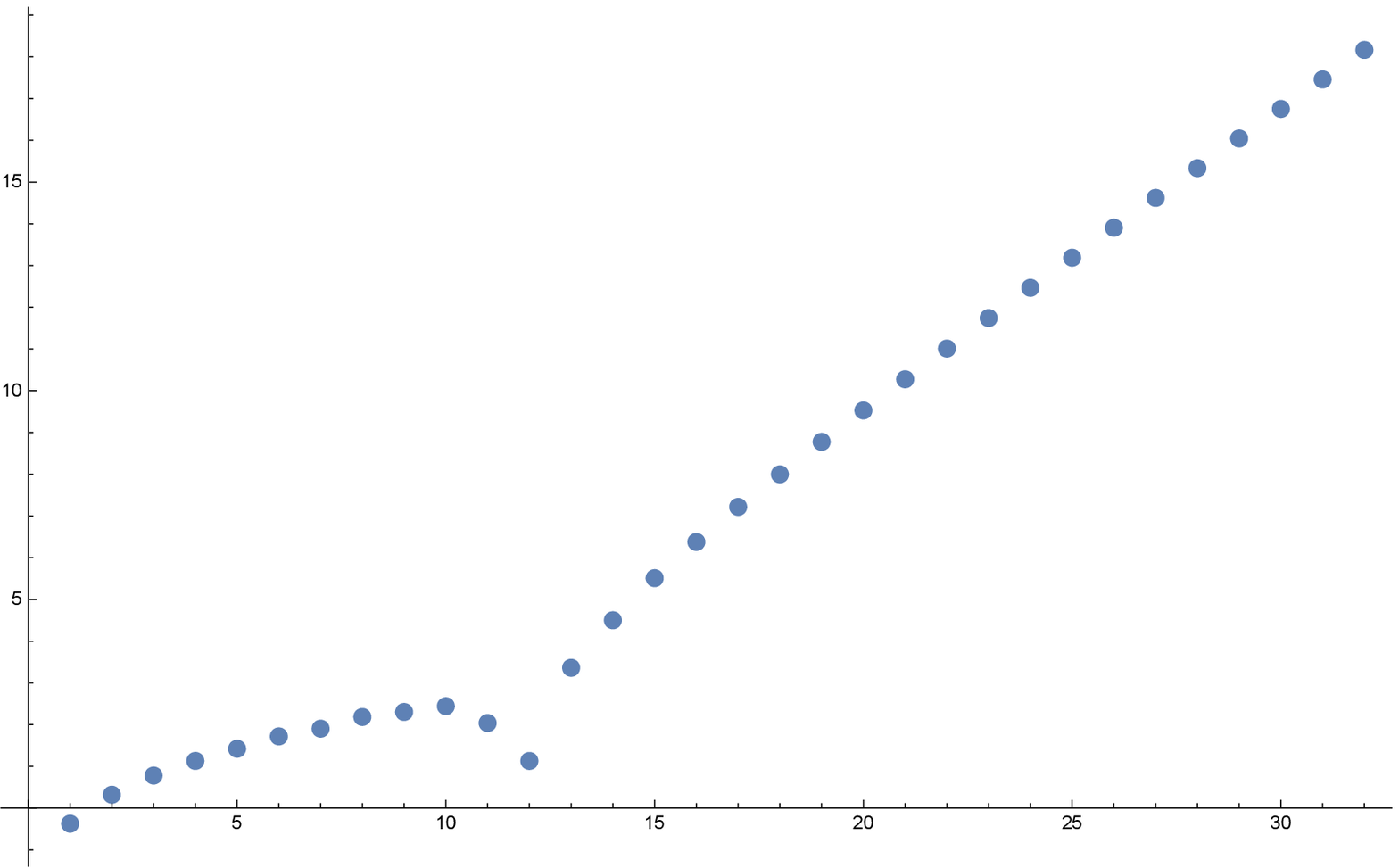}
\caption{$\log|\G^{\sst (1)\,\rm ren}_{{\rm cyc}^n}|$
from $n=1$ to 32}
\label{vacenergy 32}
\end{figure}
Extending the computation to the order $n=32$,
we find that the vacuum energy behaves 
as
$\G^{\sst (1)\,\rm ren}_{{\rm cyc}^n}\sim-C \cdot\,2^{n}$
for large $n$ (see Figure\,\ref{vacenergy 32}).
Hence, even though it is growing negatively, it would have a positive contribution:
$\sum_{n=t}^\infty \G^{\sst (1)\,\rm ren}_{{\rm cyc}^n}
\sim C \cdot 2^{t}$.
We will see in the next section that the total vacuum energy indeed
becomes a positive quantity in a different summation scheme.

\subsubsection{Zeta Function in Log Slice}\label{logslice}

When $N\to \infty$, the vacuum energy of BDYM
can organized in a different way as 
\ba\label{matrix}
\chi_{\rm\sst BDYM}(g)\eq 
\sum_{n=2}^{\infty}\,\chi_{{\rm cyc}^n}(g)
=\sum_{n=2}^{\infty}\frac1n\sum_{k|n}\,\varphi(k)\,
[\chi_{\cS^1}(g^k)]^{\frac nk}
\nn 
\eq -\chi_{\cS^1}(g)
    +\sum_{k=1}^\infty\sum_{m=1}^\infty \frac1{m\,k}\,
    \varphi(k)\,
    [\chi_{\cS^1}(g^k)]^{m}\nn 
\eq -\chi_{\cS^1}(g)+\sum_{k=1}^\infty\,\frac{\varphi(k)}{k}\,
    \chi_{\text{log},k}(g)\,,
\ea
where $\chi_{\text{log},k}$ are given by
\begin{equation}\label{logpart}
\chi_{\text{log},k}(\b,\a_1,\a_2)=
-\log\left[1-\chi_{\cS_1}\!\left(k\beta,k\alpha_1,k\alpha_2\right)\right].
\end{equation}
The above resummation of character
has been considered in 
 \cite{Sundborg:1999ue,Polyakov:2001af,Aharony:2003sx}
for the calculation of CFT partition function.
The contribution of the first term in \eqref{matrix} has already been computed in \eqref{vacsingl}, hence we will concentrate on the second term \eqref{logpart}. Before doing so, we would like to remind the reader of an argument made in \cite{Bae:2016rgm}. In principle, one may wish to consider the character obtained by \textit{first} carrying out the sum over $k$ in the second term of \eqref{matrix}. However, if $\beta_c$ is a singular point for the term \eqref{logpart} at $k=1$, then the terms with higher values of $k$ would possess singular points at $\beta_c\over k$. Further, as we increase the value of $k$, these singularities would tend to cluster around $\beta=0$ making the character highly non-analytic there. For this reason, we will compute the contribution of \eqref{logpart} to the vacuum energy, separately for a given value of $k$, and then carry out the sum over all $k$. Now given the character \eqref{logpart} and the expressions \eqref{fhndef}, we may readily evaluate the quantities
\begin{equation}
\begin{split}
f_{\text{log},k|2}(\b)&= \tfrac{1}{2}\,\sinh ^4\tfrac{\beta }{2}\, \log\!\left[
\frac{\sinh(k\,\b)\,(\cosh(k\,\b)-2)}{(\cosh(k\,\b)-1)^2}\right],\\
f_{\text{log},k|1}(\b)&= \tfrac{1}{6}\,\sinh ^2\tfrac{\beta }{2}\,(\cosh\beta-7)\, \log\! \left[
\frac{\sinh(k\,\b)\,(\cosh(k\,\b)-2)}{(\cosh(k\,\b)-1)^2}\right]\\&\qquad-\frac{k^2\,\sinh ^4\frac{\beta }{2}\,   \text{csch}^2\frac{k\,\beta}{2}}{\cosh(k\,\beta  )-2},\\
f_{\text{log},k|0}(\b)&= \log\!\left[
\frac{\sinh(k\,\b)\,(\cosh(k\,\b)-2)}{(\cosh(k\,\b)-1)^2}\right]-\frac{k^2\,\sinh ^2\frac{\beta }{2}\, (\cosh\beta-7)\,
   \text{csch}^2\frac{k\,\beta}{2}}{6\,(\cosh (k\,\beta)-2)}\\&\qquad+ \frac{k^4\,\sinh ^4\frac{\beta }{2}\left[30\,\cosh(k\,\beta)
   -5\,\cosh(2\,k\,\beta)-31\right] \text{csch}^4\frac{k\,\beta}{2}}
   {12\,(\cosh (k\,\beta)-2)^2}\,.
\end{split}
\end{equation}
On series expanding these quantities around $\beta=0$, we obtain
\begin{equation}
\begin{split}
f_{\text{log},k|2}(\b)&= -\frac{\log(-\frac{4}{k^3\,\b^3})}{32} \beta^4 - \frac{3\,k^2-\log(-\frac{4}{ k^3\,\b^3})}{192}\beta^6 +\mathcal{O}\left(\beta^7\right),\\ 
f_{\text{log},k|1}(\b)&= \frac{1-\log(-\frac{4}{k^3\,\b^3})}{4}\,\beta^2 + 
\frac{2+11\,k^2}{48}\,\beta^4 +\mathcal{O}\!\left(\beta^5\right),\\
f_{\text{log},k|0}(\b)&= -
\frac{3-\log(-\frac{4}{ k^3\,\b^3})}{2}\,\beta^2 + \frac{1-11\,k^2}{12}\,\beta^4 +\mathcal{O}\left(\beta^5\right).
\end{split}
\end{equation}
Here, one can see that precisely $\b^5, \b^3$ and $\b$
coefficients are absent in the above expansions.
Therefore, we get
\begin{equation}
\gamma_{\text{log},k|2}=0\,,\qquad \gamma_{\text{log},k|1}=0\,,\qquad \gamma_{\text{log},k|0}=0\,,
\end{equation}
and we are led to conclude that the contribution to the vacuum energy from a fixed $k$-slice vanishes. Therefore the only nontrivial contribution is from the first term and we finally find that
\begin{equation}
\Gamma^{\sst (1)\,\text{ren}}_{\text{BDYM}}= -\frac{31}{45}\,\log R\,.
\end{equation}
This suggests the shift of bulk inverse coupling
constant from $N^2-1$ to $N^2$, as we shall now show. Firstly, we have from equation \eqref{fuvdiv},
\begin{equation}
F_{\rm\sst free\, YM} = \left(N^2-1\right){31\over 45}\,\log\Lambda_{\rm\sst CFT}.
\end{equation}
Correspondingly, on the AdS side we have
\begin{equation}
\begin{split}
\Gamma_{\sst\text{BDYM}} &= \frac1g\,S_{\sst\text{BDYM}}+\Gamma^{\sst (1)\,\text{ren}}_{\sst\text{BDYM}}+\cO(g)\\
&=\frac1g\,S_{\sst\text{BDYM}} - {31\over 45}\,\log R +\mathcal{O}(g)\,.
\end{split}
\end{equation}
We are thus led to identify
\begin{equation}
\left(N^2-1\right){31\over 45}\,\log R= 
\frac1g\,S_{\sst\text{BDYM}} - {31\over 45}\,\log R\,,
\end{equation}
which suggests the identifications
\begin{equation}
    \frac1g= N^2\,,\qquad S_{\sst\text{BDYM}} = {31\over 45}\,\log R\,.
\end{equation}
Thus, as mentioned above, our computation suggests that the bulk inverse coupling constant should be shifted from $N^2-1$ to $N^2$.

\section{One-Loop Casimir Energy in Thermal AdS}
\label{sec:TAdS}

We now turn to the comparison of partition functions between the AdS$_5$ and CFT$_4$ theories, where the boundary of AdS is given by $S^1\times S^3$. 
On the AdS$_5$ side, this corresponds to taking the \emph{thermal quotient}, as described in \cite{Giombi:2008vd,David:2009xg,Gopakumar:2011qs}. 
We will now discuss the matching of 
the thermal AdS partition function
to that of CFT.
In \eqref{free F th} and \eqref{TAdS F}, 
the matching between 
$\hat F_{\rm\sst CFT}(\b)$ and $\hat\cF_{\rm\sst TAdS}(\b)$
is in fact tautological, as we review here. 
If the field in AdS$_5$ carries a  representation $\cD(\D,\bm\ell)$ of the algebra $so(2,4)$, the temperature dependent 
part of the AdS free energy is given by
(see, for example, \cite{Gibbons:2006ij})
\begin{equation}
    \hat\cF_{\cD(\D,\bm\ell)}(\beta) 
    =-\sum_{m=1}^\infty {1\over m}\,
    \chi_{\cD(\D,\bm\ell)}\!
    \left(m\,\beta\right),
\end{equation}
where $\chi_{\cD(\D,\bm\ell)}(\b)
:=\chi_{\cD(\D,\bm\ell)}(\b,0,0)$ is the blind character of $\cD(\D,\bm\ell)$\,.
In particular, this was explicitly shown in \cite{Gopakumar:2011qs,Gupta:2012he} 
for symmetric massless bosonic HS fields
by evaluating one-loop determinants.
Summing individual free energies
over the spectrum $\cH$ of the AdS theory, 
we would obtain the total free energy,
\be\label{fbeta}
    \hat\cF_{\cH}(\beta) =
    \sum_{\D,\bm\ell}
    N_{\cD(\D,\bm\ell)}\,
    \hat\cF_{\cD(\D,\bm\ell)}(\b)\,.
\ee
In the cases of the vector model dualities 
where the spectrum is simple,
the expression \eqref{fbeta} 
was evaluated to match with the order $N^0$ partition function 
$\hat F_{\rm\sst CFT}$ of the dual vector model CFT \cite{Giombi:2014yra,Shenker:2011zf,Jevicki:2014mfa}.
In the case of adjoint model CFT dualities, 
the explicit spectrum of theory is not available.
Nevertheless, using the linearity
of the free energy and character, 
the total free energy can
be directly obtained from
the total character $\chi_\cH$ as
\begin{equation}
    \hat\cF_{\cH}(\beta) 
    =-\sum_{m=1}^\infty {1\over m}\,
    \chi_{\cH}\!
    \left(m\,\beta\right).
    \label{F ch}
\end{equation}
The blind character 
$\chi_{\cH}\!\left(\beta\right)$
is nothing but the single-trace one-particle partition function of the dual CFT\footnote{This is a necessary condition for matching the spectra of the bulk and boundary theories, a requirement which we have imposed externally, in the absence of an independent construction of the bulk theory. As we have remarked previously, spectrum matching is a necessary condition for any putative AdS/CFT duality.}
and the formula \eqref{F ch} precisely defines
the multi-particle  or grand canonical partition function.
Therefore, the match of temperature 
dependent part of free energy $\cF_{\rm\sst CFT}$ and $\cF_{\rm\sst TAdS}$
is guaranteed, and we omit further discussion of this term.

Let us then turn to the term 
$\b\,\cE_{\rm\sst TAdS}$\,,
which is the effective action or free energy
of AdS$_5\simeq \mathbb R\times D^4$ with $\mathbb R\times S^3$ boundary.
With the time-translation isometry, the zeta function 
for the spectrum $\cD(\D,\bm\ell)$ in AdS$_5$ 
reduces to that of the spatial Laplacian on the disk $D^4$ as
\be
    \zeta_{\cD(\D,\bm\ell), \rm AdS_5}(z)=
    \b\,\frac{\G(z-\frac12)}{\sqrt{4\,\pi}\,\G(z)}\,\zeta_{\cD(\D,\bm\ell), D^4}(z-\tfrac12)\,.
\ee
Since the square root of the spactial Laplacian is nothing but the energy, 
the zeta function $\zeta_{\cH, D^4}$
can be directly related to the Mellin transform of the character as
\be
    \zeta_{\cD(\D,\bm\ell), D^4}(z-\tfrac12)=\tilde \chi_{\cD(\D,\bm\ell)}(z-1)\,,
    \qquad 
    \tilde\chi_{\cD(\D,\bm\ell)}(z)=\int_0^\infty d\b\,\frac{\b^{z-1}}{\G(z)}\,
    \chi_{\cD(\D,\bm\ell)}(\b)\,.
    \label{chi int}
\ee
Therefore, $\cE_{\rm\sst TAdS}$
can be interpreted as the Casimir energy in $D^4$ of the bulk theory.
Again using linearity of the expressions,
the total Casimir Energy of the  theory
having the spectrum $\cH$ 
will be given  by \cite{Giombi:2014yra}
\begin{equation}\label{casimirdef}
    \cE_{\cH} = {1\over 2}\,\tilde\chi_{\cH}(-1)\,.
\end{equation}
This Casimir energy $E_{\rm\sst TAdS}$ in the bulk will be related later on 
to $E_{\rm\sst CFT}$ of the boundary CFT, as in \cite{Giombi:2014yra}.
In order to distinguish this two quantities, we shall refer
the former/later as AdS/CFT Casimir energy.
For the evaluation of the quantity,
similarly to the $S^4$ boundary case, we can deform
the integration contour of \eqref{chi int} to Figure\,\ref{fig: ct1} to get
\be
      \tilde\chi_{\cD(\D,\bm\ell)}(z)=\frac{i}{2\,\sin(\pi\,z)}\oint_C d\b\,\frac{\b^{z-1}}{\G(z)}\,
    \chi_{\cH}(\b)\,.
\ee
If the character $\chi_{\cH}(\b)$ does not have any singularity
on the positive axis of $\b$ except for poles at $\b=0$\,,
then the contour $C$ can be shruncken to a small circle around $\b=0$ to give
\begin{equation}\label{casimirexp}
    \cE_{\cH} =-\frac12\,\oint_C \frac{d\b}{2\,\pi\,i\,\b^{2}}\,
    \chi_{\cH}(\b)\,.
\end{equation}
Hence, the AdS Casimir energy is simply given by 
$-1/2$ times of the $\b$ linear Laurent coefficient
of the character $\chi_{\cH}(\b)$\,.

\subsection{AdS Dual of Free Scalar Adjoint Model}

We now briefly return to the duality of  free adjoint scalar studied in \cite{Bae:2016rgm}. The purpose is two-fold. Firstly it will serve as a warm up for the duality of free Yang Mills which we shall soon be turning to. Secondly, it shall enable us to compare and contrast some features of the Casimir energy 
of the bulk theories dual to scalar and Yang Mills CFTs.

\subsubsection*{Casimir for A Few Orders of Regge Trajectories}

We now turn to evaluation of the AdS Casimir energy $\cE_{\rm\sst TAdS}$ for 
the first few Regge trajectory fields
of the Bulk Dual of free Adjoint Scalar (BDAS).
We shall explicitly display results for the first and second trajectories
and later plot the results for higher trajectories. 

\paragraph{Order One}

This is just the Rac representation itself. 
As discussed in \cite{Bae:2016rgm}, this 
neither belongs to the single-trace operator spectrum 
nor represent a propagating degree of freedom in the bulk. 
Nonetheless, one may formally treat it as such and compute the Casimir energy associated with it. Using the Laurent expansion of the character,
\be
    \chi_{\rm Rac}(\b)=\frac{\cosh\frac\b2}{4\,\sinh^3\frac\b2}=
    \frac{2}{\b^{2}}-\frac{\b}{120}+\cO(\b^2)\,,
\ee
we take the coefficient linear in $\b$ and multiply by $-{1\over 2}$ as per \eqref{casimirexp} to obtain
\be
    \cE_{\rm Rac}=\frac1{240}\,.
\ee

\paragraph{Order Two}

This case corresponds to the computation for the minimal Type A Vasiliev theory dual to the $O(N)$ vector model as carried out in \cite{Giombi:2014yra}. We will recover their results and review the argument for this computation indicating a shift in the bulk coupling constant from $N$ to $N-1$. We begin with the expression for the the order two character, given by
\ba
    \chi_{\rm cyc^2}(\b)\eq 
    \frac{ \sinh ^2\beta\left(\text{csch}^8\frac{\beta}{2}+16\,\coth\beta\, \text{csch}^4\beta\right)}{128} \nn
    \eq \frac{2}{\b^6}+\frac{1}{8\,\b^3}-\frac{1}{60\, \b^2}+\frac{1}{756}-\frac{\b}{120}+
    \cO\!\left(\b^2\right)\,.
\ea
Hence the Casimir energy is given by
\be
    \cE_{\rm cyc^2}=\frac1{240}\,.
\ee
We now review the argument of \cite{Giombi:2014yra} for this result indicating a shift in the bulk coupling constant. Firstly, it turns out that the total Casimir Energy in the boundary theory scales as $N$, i.e.
\begin{equation}
    F_{\sst O(N)\,\rm Scalar}(\b) = N\,\beta\, E_{\phi} + \hat F_{\sst O(N)\,\rm Scalar}(\beta)\,.
\end{equation}
Therefore, using the bulk-boundary correspondence it should be apparent that the bulk Casimir energy cannot contain loop corrections to its classical value. However, we do find a one-loop contribution. That is,
\begin{equation}
    \G_{\rm\sst  A,min} = 
    \frac1g\,S_{\rm\sst  A,min}+
    \b\,\cE_{\rm\sst  A,min}
    + \hat\cF_{\rm\sst  A,min}(\b)+\cO(\b)\,,
\end{equation}
where $S_{\sst\rm A,min}$ is the on-shell classical action.
With $\hat F_{\sst O(N)\,\rm Scalar}(\beta)=\hat\cF_{\rm\sst  A,min}(\b)$\,,
the two statements may be reconciled by requiring that
\begin{equation}
    \frac1g = N-1\,, \qquad S_{\sst\rm A,min} = {1\over 240}\,\b\,.
\end{equation}
This is the same shift as found by taking the boundary to be $S^4$ \cite{Giombi:2014iua}.

\paragraph{Order Three}

The order three character is given by
\ba
    \chi_{\rm cyc^3}(\b)\eq 
  \frac{\sinh^3\b\,\text{csch}^{12}\frac{\b}{2}+128\,\sinh(3\b)\, \text{csch}^4\frac{3\,\b}{2}}{1536}
  \nn
    \eq \frac{8}{3\,\beta^9}-\frac{1}{30\,\beta ^5}+\frac{59}{1134 \,\beta^3}-\frac{2081\,\beta}{124740}+\cO(\b^2)\,,
\ea
hence the Casimir energy is
\be
    \cE_{\rm cyc^3}=
    \frac{2081}{249480}\,.
\ee

\paragraph{Higher Orders}

\begin{figure}
\begin{center}
\includegraphics[width=0.8\textwidth]{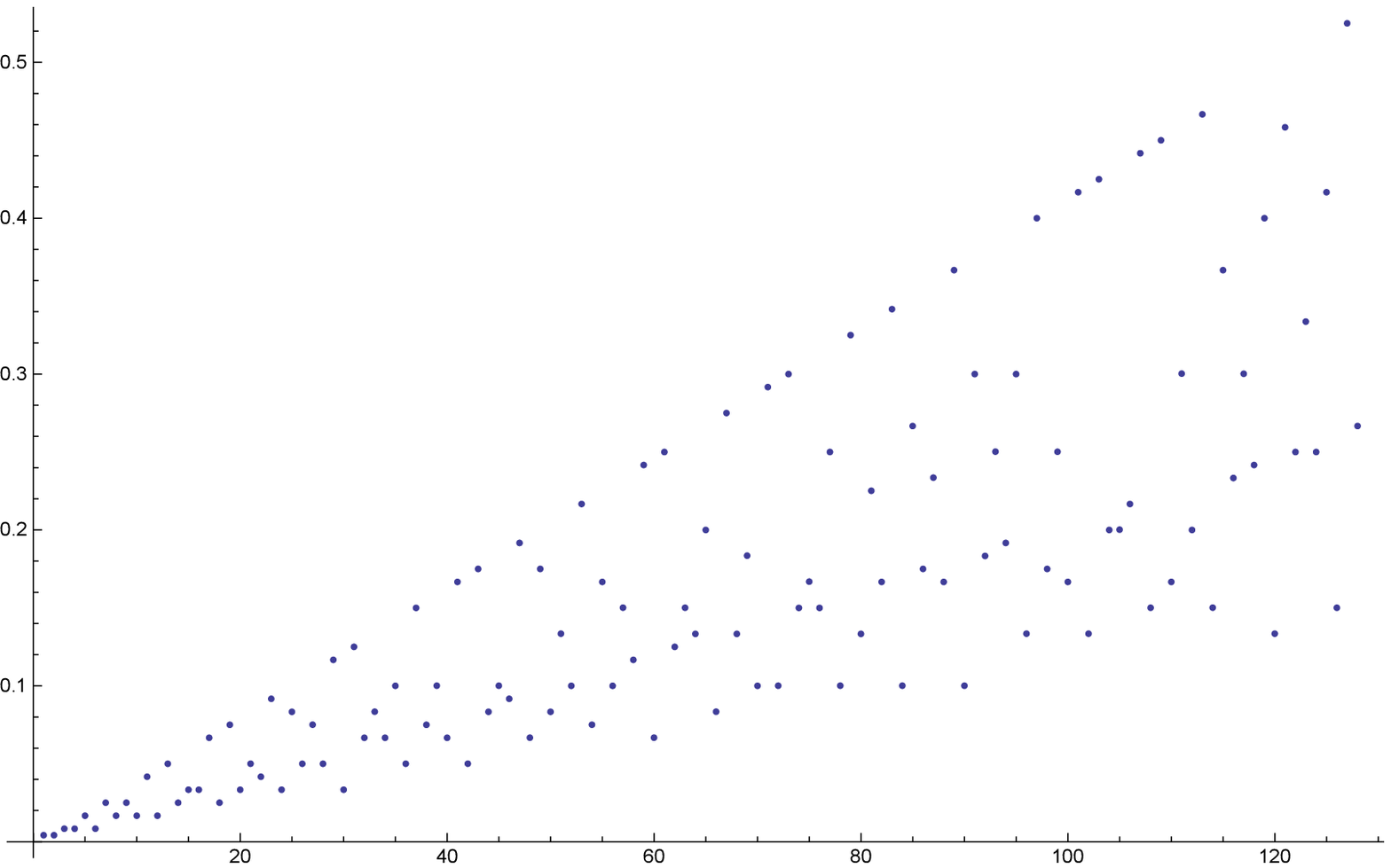}
\caption{Plot of $\cE_{\sst {\rm cyc}^n}$
for $n=1,\ldots,128$}
\label{fig: scalar Casimir}
\end{center}
\end{figure}

\begin{figure}
\begin{center}
\includegraphics[width=0.8\textwidth]{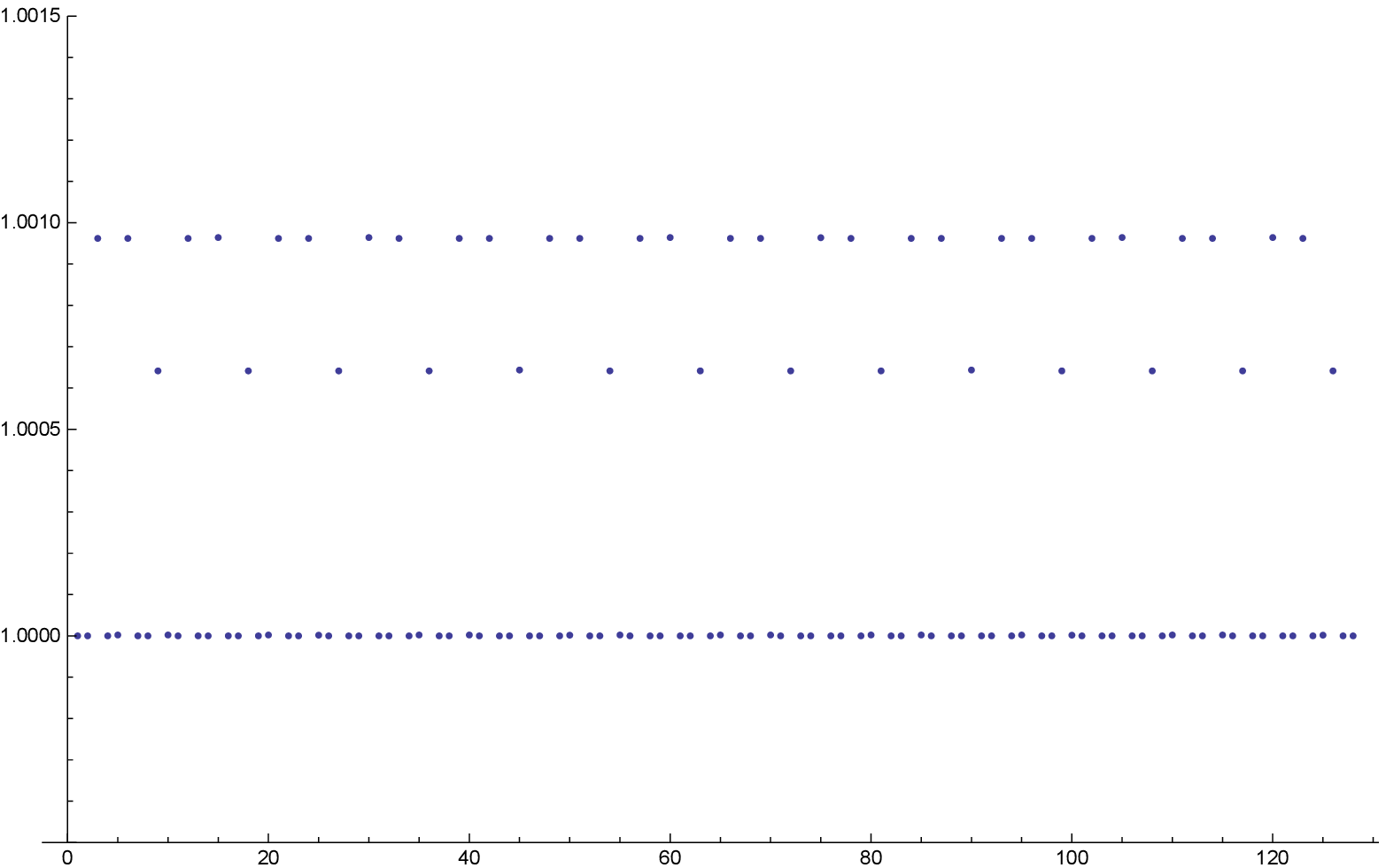}
\caption{Plot of $\cE_{\sst {\rm cyc}^n}/(\varphi(n)\,\cE_{\sst \rm Rac})$
for $n=1,\ldots,128$}
\label{scalarcasimirEuler}
\end{center}
\end{figure}

We proceed to higher orders and display the result
in Figure\,\ref{fig: scalar Casimir}.
This result exhibits an interesting pattern
analogous to what we have already reported in our previous paper \cite{Bae:2016rgm}: the values of vacuum energy (that is, the free energy in AdS$_5$ with $S^4$ boundary) become extremely simple when the order of trajectory 
is given by $n =2^k$. This pattern persists even in the Casimir energy: for $k = 1, 2,\ldots, 7$, the corresponding Casimir energy is given by $2^k/120$.
Another analogous feature to the vacuum energy case is 
that the pattern of Casimir Energy with respect to the order $n$ is compatible with the Euler totient function $\varphi(n)$. The ratio of Casimir Energy to the Euler totient function is displayed in 
Figure\,\ref{scalarcasimirEuler}.
When  $n=2^k$, the ratio $\cE_{\sst {\rm cyc}^n}/(\varphi(n)\,\cE_{\sst \rm Rac})$ becomes exactly 1. Apart from $n=2^k$, the ratio  is not exactly one, however they are very close to 
one with very small fluctuations. This behavior was similarly observed in the vacuum energy case \cite{Bae:2016rgm}. 
What is more interesting in the Casimir energy case is that the fluctuation itself has a certain pattern.
There are three group of trajectories
(consisting of three lines in  Figure\,\ref{scalarcasimirEuler}
and appearing periodically)
where the quantities $\cE_{\sst {\rm cyc}^n}/(\varphi(n)\,\cE_{\sst \rm Rac})$ are very similar with little fluctuations.
It would be  interesting to understand the reason for all these intriguing patterns, but
it is beyond the scope of the current paper.

\subsubsection*{Casimir in Log Slice}

As illustrated in the previous section, the AdS Casimir energy grows rapidly with respect to the order of Regge trajectory $n$. Therefore, 
the total Casimir energy of BDAS requires suitable regularization process as the vacuum energy. 
Instead, we organize the total Casimir energy of BDAS
in a different way by making use of the resummation formula of character as in \eqref{matrix}. The Mellin transform of the character
can be also cast in the same resummation form as
\ba
    \tilde\chi_{\rm\sst BDAS}(z) = -\tilde\chi_{\rm\sst  Rac}(z)
     + \sum_{k=1}^{\infty} \frac{\varphi(k)}{k}
     \,\tilde\chi_{{\rm log},k}(z)\,,
    \label{Casimir_adj}
\ea
where $\chi_{{\rm log},k}(\b)=-\log[1-\chi_{\rm Rac}(k\,\b)]$\,.
In this section, we consider even the arbitrary dimensional case,
that is, for the Bulk Dual of free Adjoint Scalar (BDAS) on $S^{1}\times S^{d-1}$\,.
The character of Rac is given in any $d$ as
\be
	\chi_{\rm\sst  Rac}(q)=\frac{q^{\frac{d-2}2}(1+q)}{(1-q)^{d-1}}\qquad
	\text{or}\qquad
    \chi_{\rm\sst  Rac}(\b)=2^{2-d}\,\frac{\cosh\frac\b2}{(\sinh\frac\b2)^{d-1}}
\ee
Here, the variable $q$ is defined by $e^{-\b}$.

Firstly we focus on the log piece in \eqref{Casimir_adj}.
Here we have two ways to proceed. 
One way is the expansion of the character,
\be
    \chi_{\log,k}(\b)
    =-\log[1-\chi_{\rm\sst  Rac}(k\,\b)]
    =3\,\ln\b-\log\!\left(-\frac{2}{k^2}\right)
    +\frac{k^3\,\b^3}{2}+\cO(\b^4)\,,
\ee
ignoring the branch contribution.
Here, we can see that the above character does not have
any linear term in $\b$ hence the corresponding 
Casimir energy vanishes.
The same conclusion can be drawn in a different way:
we first perform the change of variable $k\,\b \rightarrow \b$
inside the Mellin integral which allows to factor 
out the $k$ dependence as
\be
	\tilde\chi_{{\rm log},k}(z)
	=-\int_0^\infty d\b\,\frac{\b^{z-1}}{\G(z)}\,
	\log[1-\chi_{\rm\sst Rac}(k\,\b)]
	=k^{-z}\,\tilde\chi_{{\rm log},1}(z)\,.
\ee	
The $k$ summation can be independently
carried out from $\tilde\chi_{{\rm log},1}(z)$ as
\be
    \sum_{k=1}^{\infty}\,\frac{\varphi(k)}{k}\,	\tilde\chi_{{\rm log},k}(z)
    =\left(\sum_{k=1}^{\infty}\,\frac{\varphi(k)}{k^{z+1}}\right)    \tilde\chi_{{\rm log},1}(z)
    =\frac{\zeta(z)}{\zeta(z+1)}\,\tilde\chi_{{\rm log},1}(z)\,.
    \label{k sum done}
\ee
For the remaining part  $\tilde\chi_{{\rm log},1}(z)$,
instead of deforming the contour, we 
perform an integration by part to get
\be
	 \tilde\chi_{{\rm log},1}(z)=
	 -\int_{0}^{\infty} d\b\,\frac{\b^{z-1}}{\Gamma(z)}\,
	\log\left[1-\chi_{\rm Rac}(\b)\right]
	=-\int_{0}^{\infty}d\b\,\,\frac{\b^{z}}{\Gamma(z+1)}\,
	\frac{\partial_{\b}\chi_{\rm Rac}(\b)}{1-\chi_{\rm Rac}(\b)}+i\,c\,,
\ee
where $c$ is a constant arising from
the singularity of logarithmic function.
Let us first ignore this imaginary constant piece.
In the rest of integral, the character  
enters through the form,
\be
	\frac{q\,\partial_{q}\,\chi_{\rm\sst Rac}(q)}{\chi_{\rm\sst Rac}(q)-1}=
	q^{\frac d2-1}\,
	\frac{1+  q^2 - \frac d2\, (1 + q)^2}{
(1 - q)^d - q^{\frac d2-1} (1 - q^2)}
	=q^{\frac d2-1}\left(\frac{d-1}{1-q}+R(q)\right).
\ee
The quantity have the simple pole at $q=1$ and $R(q)$ is 
the regular rational function in the real axis of $q$.
Therefore, this function can be decomposed as
\be
    R(q)=\sum_{n=1}^{d-1}\,\frac{a_{n}}{1-b_{n}\,q}\,,
    \label{R q}
\ee
with proper complex coefficients $a_{n}$ and $b_{n}$\,. This decomposition allows the analytic $\b$-integral by Hurwitz-Lerch transcendent $\Phi\left(z,s,a\right)$.
\be
	\tilde\chi_{{\rm log},1}(z)
	=-(d-1)\,\zeta\left(z+1,\frac{d-2}2\right)
	-\sum_{n=1}^{d-1} a_{n}\,\Phi\left(b_{n},z+1,\frac{d-2}2\right)\,.
	\label{log 1}
\ee
Putting $z=-1$ with using properties of $\Phi(z,s,a)$, we get 
\be
	\tilde\chi_{{\rm log},1}(-1)=-(d-1)\,\zeta\left(0,\frac{d-2}2\right)-
	R(1)\,.
\ee
Finally, using the information,
\be
    \zeta\!\left(0,\frac{d-2}2\right)=\frac{3-d}2 \,,
    \qquad 
    R(1)=\frac{(d-1)(d-3)}2\,,
\ee
we find that two terms in \eqref{log 1} precisely cancel to each other
hence $\tilde\chi_{\log,k}(-1)=0$\,.

Since the log pieces in \eqref{Casimir_adj} do not contribute to the 
AdS Casimir energy, the nontrivial contribution totally come from the first term in \eqref{Casimir_adj}.
For its evaluation, we need to extract the $\b$ linear coefficient
of the $\chi_{\rm Rac}(\b)$\,.
Analytic expression involves multiple summation, hence let us provide
the result for the first few dimensions. First we note that in all odd $d$, 
the function $\chi_{\rm Rac}(\b)$ is even and the linear term is absent.
About the even $d=2,4,6,8,10,12$, the total Casimir energy for BDAS 
is given by
\be
    \cE_{\rm\sst BDAS}=\frac1{12}\,,
    \quad -\frac1{240}\,, 
    \quad \frac{31}{60480}\,,
     \quad -\frac{289}{3628800}\,,
      \quad \frac{317}{22809600}\,,
       \quad -\frac{6803477}{2615348736000}\,.
\ee
The sign of Casimir energy is hence positive in $d=4n-2$ dimensions
whereas negative in $d=4\,n$ and they fastly decrease as $d$ increases.
We now examine the implication of this result on the dictionary between the boundary parameter $\mathsf{N}=N^2-1$ and the bulk coupling constant $g$. Firstly, the total Casimir energy in the boundary theory scales as $N^2-1$, i.e.
\begin{equation}
    F_{\rm\sst Adj\,Scalar}(\b) = \left(N^2-1\right)\beta\,E_\phi + 
    \hat F_{\rm\sst Adj\,Scalar}(\beta)\,,
\end{equation}
where the boundary scalar Casimir energy is given by
\begin{equation}
    E_\phi = \cE_{\text{Rac}} = {1\over 240}\,.
\end{equation}
Now the bulk Casimir energy does contain a one-loop contribution. That is,
\begin{equation}
    \G_{\rm\sst BDAS}(\b) = \frac1g\,S_{\rm\sst BDAS}
    + \b\,\cE_{\rm\sst BDAS}+
    \hat\cF_{\rm\sst BDAS}(\b)+\cO(g)\,.
\end{equation}
Again, with the match between $\hat F_{\rm\sst Adj\,Scalar}(\beta)$
and $\hat\cF_{\rm\sst BDAS}(\b)$\,,
the correspondence holds if we impose
\begin{equation}
    \frac1g = N^2\,, \qquad S_{\rm\sst BDAS} = {1\over 240}\,\b\,.
\end{equation}
This is the same shift found to match the 
total vacuum energy of BDAS to the boundary free adjoint scalar CFT \cite{Bae:2016rgm}.

\subsection{AdS Dual of Free Yang-Mills}

With the experience of BDAS, we now turn to the main computation of this section, 
the total Casimir energy of the Bulk theory in AdS$_5$ Dual 
to the free Yang-Mills (BDYM) on the boundary $S^1\times S^3$\,.
This task has been carried out in \cite{Basar:2014hda}.
After presenting independent analysis of the current work,
we shall comment more on its relation to the last reference.

\subsubsection*{Casimir Energy for First Few Regge Trajectories}

 We first do an order-by-order computation, 
 evaluating the Casimir energies
 of the fields in first few Regge trajectories of BDYM.
 Let us remind that order $n$ trajectory fields
 are dual to the CFT operators involving $n$-th power of $\bm F_{ab}$
 hence correspond to $n$-ple cyclic tensor product of  $\cS_1$\,.
 Here, we show the results as $n$ runs from one to 16, and then finally the computation for the full theory. 
 
\paragraph{Order One}

The relevant character is that of spin-one singleton,
\be 
    \chi_{\cS^1}(q)
    = \frac{2\,q^2 \,(3-q) }{(1-q)^3}
    =\frac{4}{\b^3}-\frac{2}{\b}+1-\frac{11 \,\b}{60}+O(\b^2)\,.
    \label{one S1}
\ee 
Taking the coefficient of $\b$-linear term, we find
\be
    \cE_{\cS^1}=\frac{11}{120}\,.
    \label{E S1}
\ee 

\paragraph{Order Two}

The character for the first Regge trajectory is 
\be 
    \chi_{\rm cyc^2}(\b)
    = \frac{8}{\b^6}-\frac{8}{\b^4}+\frac{17}{4 \b^3}+\frac{19}{15 \b^2}-\frac{5}{2 \b}+\frac{1307}{945}-\frac{11 \b}{30}+\cO\!\left(\b^2\right)
\ee 
hence we obtain
\be
    \cE_{\rm cyc^2}=\frac{11}{60}\,.
\ee 
This reproduces the  Casimir energy for the minimal Type C bulk theory 
(see the equation (4.2) of \cite{Beccaria:2014zma}) simply because the field content coincides. We now digress briefly to review the proposal of \cite{Beccaria:2014zma} for how this result indicates the same shift in the dictionary between the bulk dimensionless inverse coupling constant and the boundary parameter $\mathsf{N}$, which is simply $N$ for the fundamental representation of $O(N)$. Firstly, the boundary Casimir energy turns out to be proportional to $N$, i.e. 
\begin{equation}
    F_{\sst O(N)\,\rm Maxwell}(\b) = N\,\beta\,E_V + 
    \hat F_{\sst O(N)\,\rm Maxwell}(\beta)\,.
\end{equation}
Therefore, using the bulk-boundary correspondence it should be apparent that the bulk Casimir energy cannot contain loop corrections to its classical value. However, we do find a one-loop contribution. That is,
\begin{equation}
    \G_{\rm\sst C,min} = \frac1g\,S_{\rm\sst C,min}
    +\cE_{\rm\sst C,min}+
    \hat\cF_{\rm\sst C,min}(\b)+\cO(g)\,,
\end{equation}
where $S_{\rm\sst C,min}$ is the on-shell classical action. 
The correspondence would work by requiring 
\begin{equation}
    \frac1g = N-2\,, \qquad S_{\rm\sst C,min}=\frac{11}{60}\,\b\,,
\end{equation}
together with the match $\hat F_{\sst O(N)\,\rm Maxwell}(\beta)=\hat\cF_{\rm\sst C,min}(\b)$\,.
This is the same shift suggested by the vacuum energy computation.

\paragraph{Order Three}

The order three Regge trajectories contain only massive HS fields
including Goldstone modes for massless HS.
The relevant character here is 
\be 
    \chi_{\rm cyc^3}(\b)
    =\frac{64}{3\,\b^9}-\frac{32}{\b^7}+\frac{16}{\b^6}+\frac{196}{15\,\b^5}-\frac{16}{\b^4}+\frac{12422}{2835\, \b^3}+\frac{38}{15 \,\b^2}-\frac{26}{9\,\b}+\frac{1354}{945}
    -\frac{18767\,\b}{44550}+\cO(\b^2)\,,
\ee 
hence we get the AdS Casimir energy,
\be
    \cE_{\rm cyc^3}=\frac{18767}{89100}\,.
\ee 

\paragraph{Higher Orders}

\begin{figure}[h]
\centering
\begin{minipage}{0.48\textwidth}
\centering
\includegraphics[width=\textwidth]{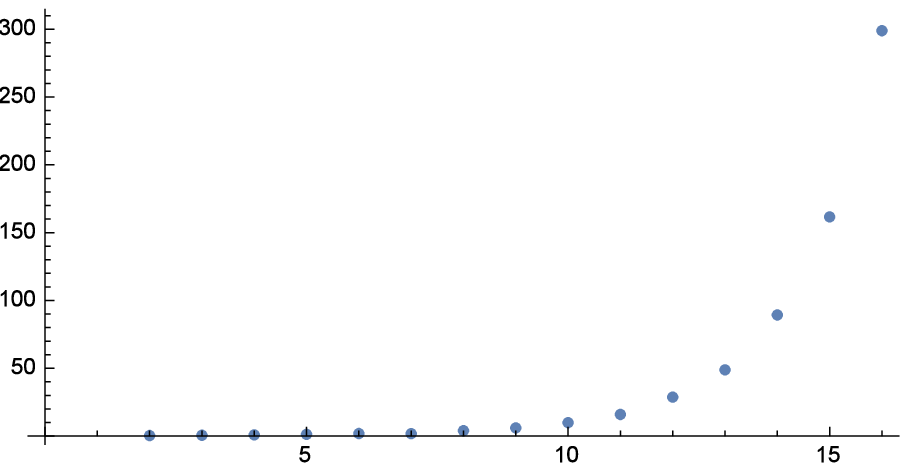}
\caption{Plot of $\cE_{{\rm cyc}^n}$ up to $n=16$}
\label{vectorcasimir}
\end{minipage}\quad
\begin{minipage}{0.41\textwidth}
\centering
\includegraphics[width=\textwidth]{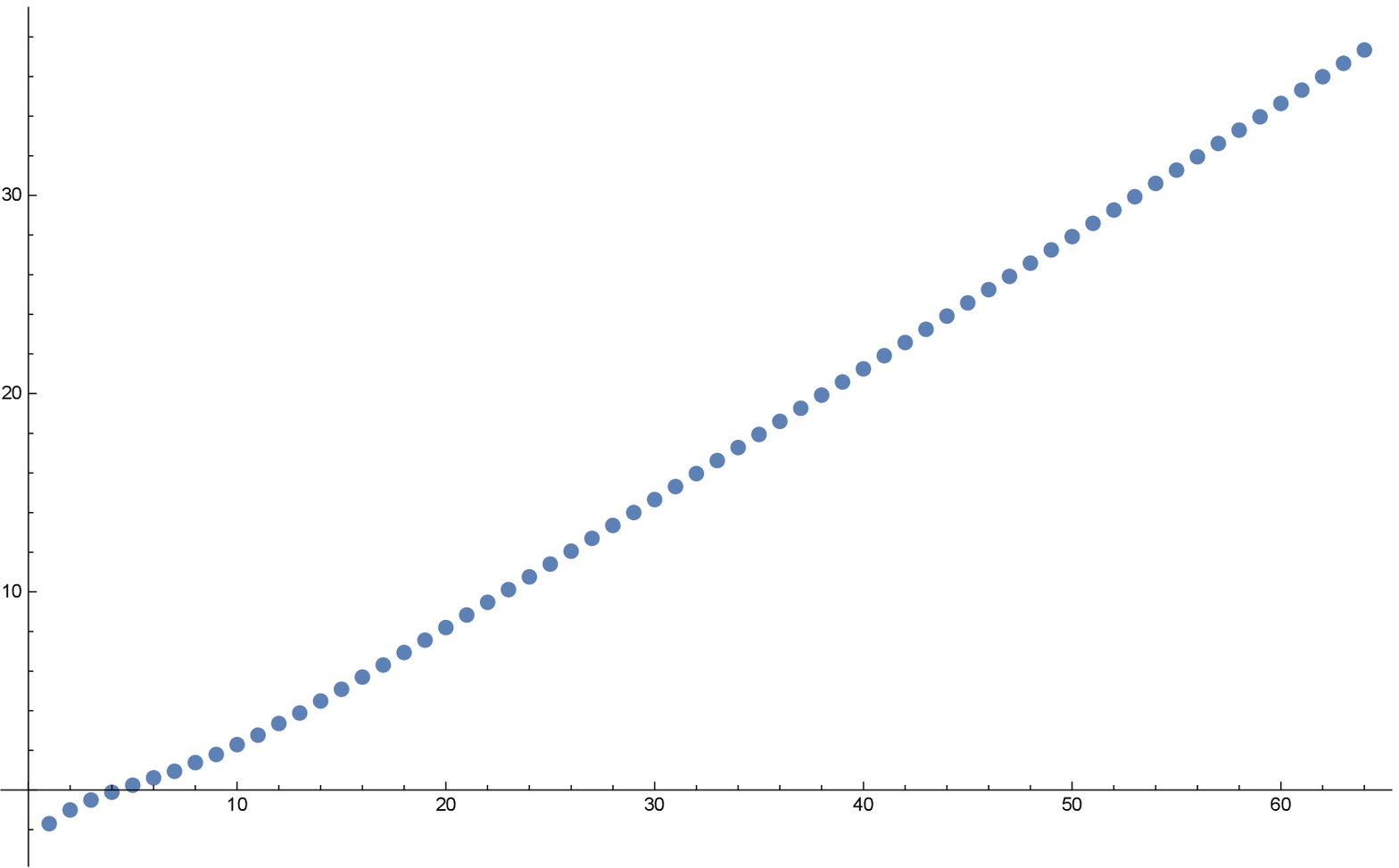}
\caption{$\log(\cE_{{\rm cyc}^n})$
from $n=1$ to 32}
\label{Casimir 32}
\end{minipage}
\end{figure}

For the free Yang-Mills case in $d=4$,
the numerical values of Casimir energy up to order 16 are displayed in the Figure \ref{vectorcasimir},
which shows an exponential growth.
Extending the computation to $n=32$ and taking log, 
we find that the asymptotic behavior is again $ \cE_{\rm cyc^n} \sim C\,2^n$.
This does not shows any chaotic pattern like BDAS. 
Moreover, in contrast with the vacuum energy,
the Casimir energy does not exhibit sign flip at least up to $n=32$\,. 

\subsubsection*{Casimir in Log Slice}

The AdS Casimir energies for the first few Regge trajectories 
exhibit an exponential-like growth with respect to the order of Regge trajectory $n$, hence there sum would require a regularization.
We bypass this issue by slicing the total Casimir energy in a different as
\be
    \tilde\chi_{\rm\sst BDYM}(z) = -\tilde\chi_{\cS^1}(z)
     + \sum_{k=1}^{\infty} \frac{\varphi(k)}{k}
     \,\tilde\chi_{{\rm log},k}(z)\,,
   \label{Casimir fYM}
\ee
where $\chi_{{\rm log},k}$ is given in \eqref{logpart}.
The first term computation is 
already carried out in \eqref{one S1} and \eqref{E S1}.
To evaluate second term, we again have two options.
First option is to expand the character ignoring the branch cut contribution as
\be 
    \chi_{\log,k}(\b)
    =-\log\!\left(-\frac4{k^3\,\b^3}\right)
    +\frac{k^2\,\b^2}2+\cO(\b^3)\,.
\ee
The absence of the linear term in $\b$ shows that
the log pieces do not contribute to the Casimir energy.
In the second option, we first change 
the integration variable as $k\,\b \rightarrow \b$ to get
\ba
    \sum_{k=1}^{\infty} \frac{\varphi(k)}{k}
     \,\tilde\chi_{{\rm log},k}(z)
	=\frac{\zeta(z)}{\zeta(z+1)}\,\tilde\chi_{{\rm log},1}(z)\,,
\ea
Like the case of BDAS, we evaluate the remaining part $\tilde\chi_{{\rm log},1}(z)$ 
by an integration by part as
\be
	\tilde\chi_{{\rm log},1}(z)
	=-\int_{0}^{\infty}d\b\,\frac{\b^{z}}{\Gamma(z+1)}\,
	\frac{q\,\partial_{q}\,\chi_{\cS^1}(q)}{\chi_{\cS^1}(q)-1}
	+i\,c\,,
\ee
where $c$ is the contribution from the log singularity and 
the rest can be recast into
\be
	\frac{q\,\partial_{q}\,\chi_{\cS^1}(q)}{\chi_{\cS^1}(q)-1}= \frac{3}{1-q}+R(q)\,,
	\qquad
	R(q)=\frac{3\,(-1+2\,q+q^2)}{(1+q)(1-4\,q+q^2)}\,.
\ee
This decomposition allows an analytic $\b$-integral via Hurwitz zeta function and Hurwitz Lerch transcendent analogously to previous section.
Following the same line of calculation, we arrive
\be 
	\tilde\chi_{{\rm log},1}(z) = -3 \, \zeta(0,0) - R(1)\,.
\ee 
Since \mt{\zeta(0,0)=1/2} and \mt{R(1)=-3/2},
the two terms in the above equation exactly cancel each others
and  this proves the vanishment of log contribution.
Let us comment here on the relation of the above calculation
to that of \cite{Basar:2014hda}.
In the latter paper, the author calculated the Casimir energy through the formula
\be
    \cE_{\cH} =\lim_{q\to 1}\frac12\,q\,\partial_q\,\chi_{\cH}(q)=\lim_{\b\to 0} -\frac12\,\partial_\b \chi_{\cH}(\b)\,.
\ee
From the expression after the second equality,  
it clearly gives the $\b$ linear term
in $\chi_{\cH}$ up to the singular terms in the $\b\to0$ limit. 
Therefore, one may consider this as small $\b$ cut-off regularization 
whereas the current paper makes use of the zeta function regularization. 
Despite of the difference in the regularization methods, 
both analysis give the same answers.

In conclusion, the total Casimir energy receives the contribution 
only from the first term of \eqref{Casimir fYM}\footnote{In \cite{Basar:2014hda},
only the log part was considered without the first term in \eqref{Casimir fYM},
hence the authors found vanishing Casimir energy.} hence it
is given by
\be
    \cE_{\rm\sst BDYM} = -{11\over 120}\,.
\ee
We now examine the implication of this result on the dictionary between the boundary parameter $\mathsf{N}=N^2-1$ and the bulk coupling constant $g$. Firstly, the total Casimir energy in the boundary theory scales as $N^2-1$, i.e.
\begin{equation}
    F_{\sst\rm free\,YM}(\b) = \left(N^2-1\right)\,\beta\, E_V + 
    \hat F_{\sst\rm free\,YM}(\b)\,
\end{equation}
where the boundary spin-one Casimir energy is given by
\begin{equation}
    E_V = \cE_{\cS^1} = {11\over 120}\,.
\end{equation}
Now the bulk Casimir energy does contain a one-loop contribution. That is,
\begin{equation}
    \G_{\rm\sst BDYM} = 
    \frac1g\,S_{\rm\sst BDYM}
    + \b\,\cE_{\rm\sst BDYM}
    +\hat\cF_{\rm\sst BDYM}(\b)+\cO(g)\,,
\end{equation}
where $S_{\rm\sst BDYM}$ is the on-shell classical action. 
The two statements may be reconciled by requiring that
\begin{equation}
    \frac1g = N^2\,, \qquad S_{\rm\sst BDYM} = {11\over 120}\,\b\,,
\end{equation}
which is the same shift as found for the vacuum energy computation in Section \ref{logslice}.

\section{Conclusion}
\label{conclusion}

In this paper, we have examined 
the AdS/CFT of free Yang-Mills with gauge group $SU(N)$ in the large $N$ limit.
The Bulk Dual theory of free Yang-Mills (BDYM) is not given independently
but defined through the holographic correspondence. 
Its field content is dictated by the single-trace operators of free Yang-Mills
and their interactions are to reproduce 
the correlation functions of the corresponding 
CFT operators
through Witten diagrams. 
Based on this, we characterized and discussed various properties of BDYM.
Once the bulk classical theory is determined from CFT,
the quantum property of the former should be consistent 
with the $N$ dependence of the CFT.
Let us remark that non-triviality of the duality 
start to appear from this point when the bulk theory is not independently defined.
Here, we considered the simplest  quantum property of BDYM ---
the one-loop partition function. Depending on the 
background of the bulk geometry, either AdS$_5$ with $S^4$ boundary
or thermal AdS$_5$ with $S^1\times S^3$ boundary,
the partition function gives vacuum or Casimir energy of the bulk theory.
These quantities are defined as the sum of individual vacuum/Casimir energies
of infinitely many fields in the theory. 
The resummations of these quantities were carried out by utilizing
the method of Character Integral Representation of Zeta function (CIRZ) 
introduced in \cite{Bae:2016rgm}.
A main difference between the vector and adjoint model dualities
is that 
the corresponding AdS theory of adjoint model has \emph{infinite times more} fields 
than that of vector models. By employing analogy with string theory,
vector model operator spectrum only consists of a single Regge trajectory
which itself contains infinitely many fields,
whereas that of adjoint model involves infinitely many trajectories.
On the one hand, the infinity of fields in a Regge trajectory can be simply regularized
by employing zeta function: the variable $z$ regularizes both
of UV divergence and the infinity of fields (the former may be regarded as the infinity of oscillator modes).
On the other hand, the infinity of trajectories is not controlled by the regulator $z$ and one might need to introduce another regulator for this.
However, 
when $N\to\infty$, the total vacuum/Casimir energy can be rearranged such
that it is given by a different sum.
In this new summation formula, we showed using CIRZ that the summand vanishes except for one term. This result is obtained up to one technical subtlety,
which will be discussed in the following paragraph.
The result suggests that the bulk coupling constant should be identified to $1/N^2$
rather than $1/(N^2-1)$\,,
and the on-shell evaluation of the BDYM classical action
is $\frac{31}{45}\,\log R$ and  $\frac{11}{120}\,\b$
for the background AdS and thermal AdS, respectively.
Besides the holography of free Yang-Mills,
we have also revisited the type-C theory dual to Maxwell vector models and the holography of free adjoint scalar CFT.

\begin{figure}[h]
\centering
\begin{minipage}{0.4\textwidth}
\centering
\includegraphics[width=0.9\linewidth]{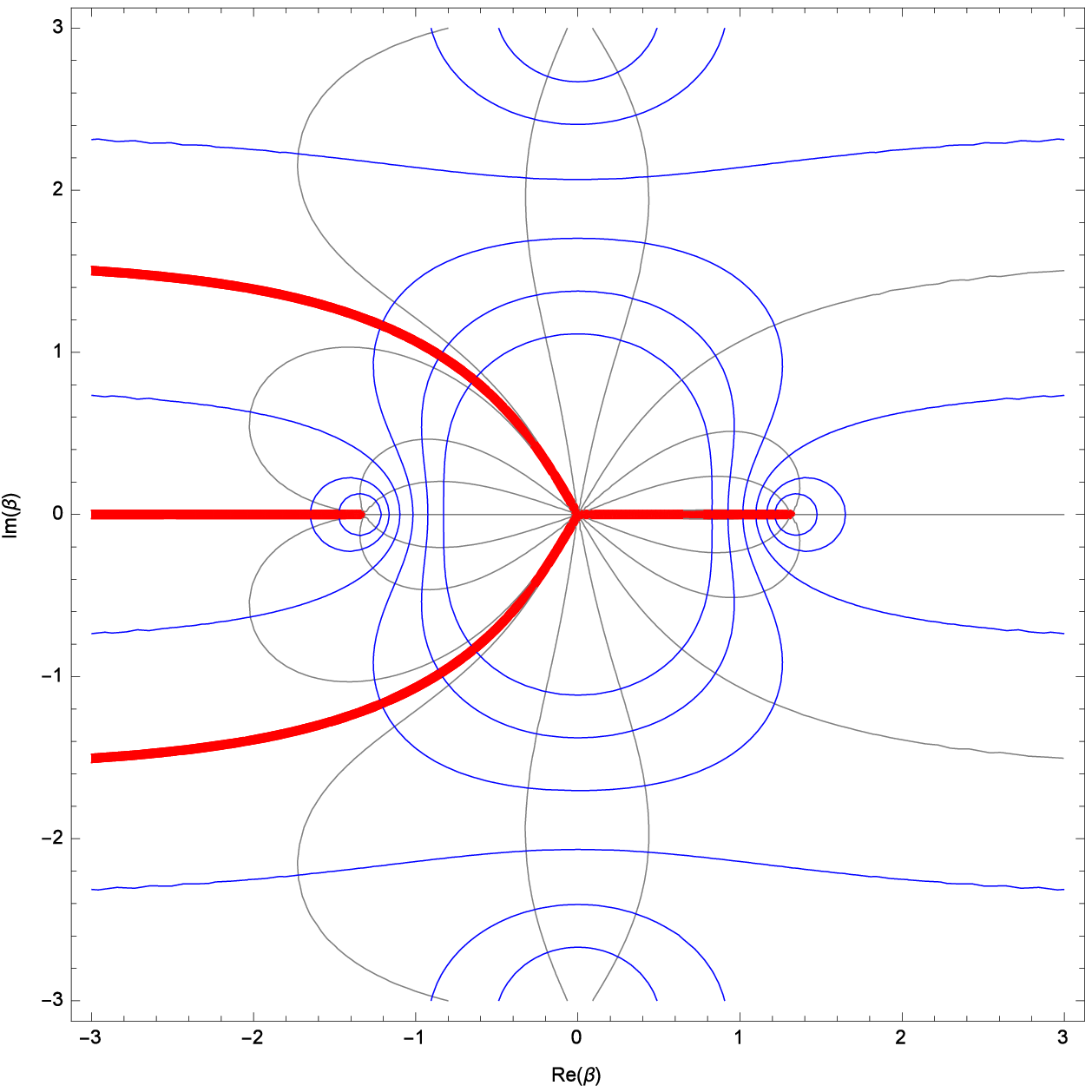}
\caption{Branch cut of the character in the Bulk Dual theory of free Yang-Mills}
\label{BDYM branch}
\end{minipage}\qquad
\begin{minipage}{0.4\textwidth}
\centering
\includegraphics[width=0.9\linewidth]{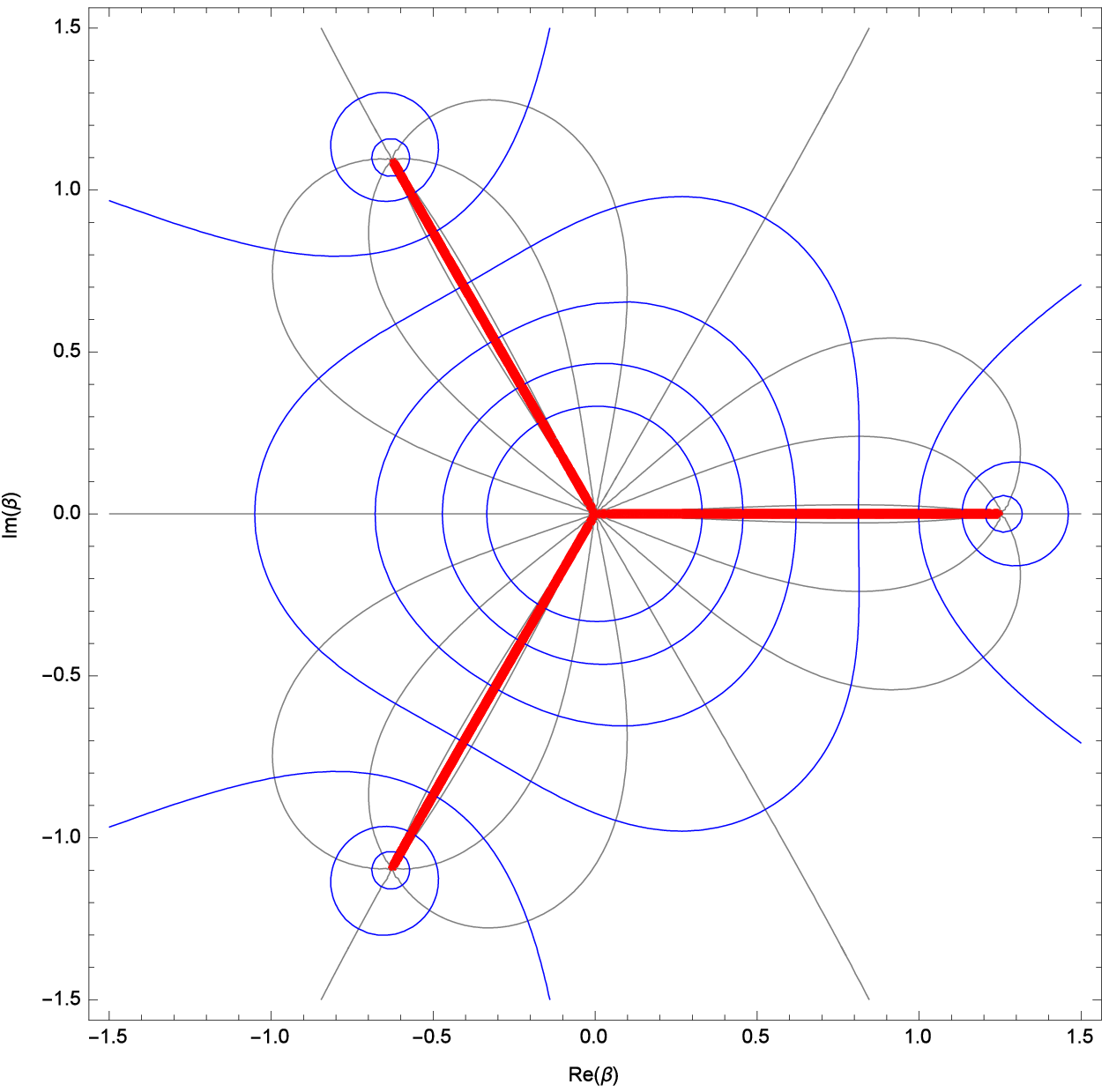}
\caption{Branch cut of the character in the Bulk Dual theory of Adjoint Scalar}
\label{BDAS branch}
\end{minipage}
\end{figure}

Let us conclude this paper with discussions on the above-mentioned subtlety.
In Section \ref{logslice}, we have shown that the log piece contributions
all vanish as the corresponding characters do not have the relevant
$\b^{2n+1}$ terms. 
As discussed in Section \ref{cirz} below \eqref{vac ener},
there is another way to proceed for the vacuum energy
is to evaluate the contour integral for each of this log pieces.
As discussed in \cite{Bae:2016rgm}, the difference between two method 
is whether we include the branch cut contributions of  the contour integral
of log pieces. 
In the  vacuum energy computation in the background of AdS$_5$ with $S^4$ boundary,
these contributions have very nontrivial $k$ dependence, where $k$ is the summation
variable, hence it is hard to encompass.
However, in the Casimir energy computation in the background of AdS$_5$ with $S^1\times S^3$ boundary, the $k$ dependence nicely factorizes in \eqref{k sum done}
to give $\zeta(z)/\zeta(z+1)$
which gives the constant factor $\zeta(-1)/\zeta(0)=\frac16$ in the \mt{z=-1} limit.
Therefore, one may expect to evaluate the 
contour integral,
\be
    \oint_C\,\frac{d\b}{2\,\pi\,i\,\b^2}\,
    \log[1- \chi_{\cS^1}(\b)]
    =\oint_C\,\frac{d\b}{2\,\pi\,i\,\b^2}\,
    \log\!\left[\frac{\sinh\b\,(\cosh\b-2)}{(\cosh\b-1)^2}\right],
\ee
where $C$ is the contour encircling 
the singularity of the integrand that is connected to the origin.
It turns out that the branch cut
of the free Yang-Mill's integrand (Figure\,\ref{BDYM branch}) is not confined in a finite region differently
from the adjoint scalar case
(Figure\,\ref{BDAS branch}).
It would be very interesting 
to understand the implication of this 
branch cut contribution and 
also the difference between two models.

\acknowledgments

We are grateful to 
Eduardo Conde, Karapet Mkrtchyan and Evgeny Skvortsov 
for useful discussions. 
The work of JB is supported by the National Research Foundation of Korea grant number NRF-2015R1D1A1A01059940. 
The work of EJ was supported in part by the National Research Foundation of Korea through the grant NRF2014R1A6A3A04056670 and the Russian Science Foundation grant 14-42-00047 associated with Lebedev Institute. 
The work of SL is supported by the Marie Sklodowska Curie Individual Fellowship 2014. 
EJ and SL would also like to thank the organizers and participants of the MIAPP programme ``Higher Spin Theory and Duality" for several interesting discussions as well as hospitality when part of this work was carried out.

\bibliographystyle{JHEP}
\bibliography{matrix}
\end{document}